\documentclass[11pt]{article}
\usepackage{amssymb, latexsym, amsmath, amsthm}
\usepackage{subfig}
\usepackage{amsfonts, graphics}
\def\R{\mathbb R}
\def\N{\mathbb N}

\def\be{\begin{equation}}
\def\ee{\end{equation}}
\def\bea{\begin{eqnarray}}
\def\eea{\end{eqnarray}}
\def\beas{\begin{eqnarray*}}
\def\eeas{\end{eqnarray*}}

\newcommand{\prfe}{\hspace*{\fill} $\Box$

\smallskip \noindent}

\begin{document}
\sloppy
\newtheorem{theorem}{Theorem}[section]
\newtheorem{definition}[theorem]{Definition}
\newtheorem{proposition}[theorem]{Proposition}
\newtheorem{example}[theorem]{Example}
\newtheorem{cor}[theorem]{Corollary}
\newtheorem{lemma}[theorem]{Lemma}
\theoremstyle{remark}
\newtheorem*{remark}{Remark}

\renewcommand{\theequation}{\arabic{section}.\arabic{equation}}

\title{Mass-radius spirals for steady state families of the 
       Vlasov-Poisson system}

\author{Tobias Ramming, Gerhard Rein\\
        Fakult\"at f\"ur Mathematik, Physik und Informatik\\
        Universit\"at Bayreuth\\
        D-95440 Bayreuth, Germany\\
        email: tobias.ramming@uni-bayreuth.de\\
        \phantom{ema}gerhard.rein@uni-bayreuth.de}

\maketitle
\begin{abstract}
We consider spherically symmetric steady states
of the Vlasov-Poisson system, which describe equilibrium 
configurations of galaxies or globular clusters.
If the microscopic equation of state, i.e., the dependence
of the steady state on the particle energy (and angular momentum)
is fixed, a one-parameter family of such states is obtained.
In the polytropic case the mass of the state along such a one-parameter 
family is a monotone function of its radius. 
We prove that for the King, Woolley-Dickens, and
related models this mass-radius relation takes the form of a spiral. 
\end{abstract}
\section{Introduction}
\setcounter{equation}{0}
We consider the stationary Vlasov-Poisson system
in the gravitational case:
\be \label{vlasov}
v \cdot \nabla_x f - \nabla U \cdot \nabla _v f=0,
\ee
\be \label{poisson}
\Delta U = 4 \pi \rho,\ \lim_{|x|\to \infty} U(x)=0,
\ee
\be \label{rhodef}
\rho (x) = \int f(x,v)\,dv.
\ee
Here  $f=f(x,v)\geq 0$, a function of position $x\in \R^3$ and velocity 
$v \in \R^3$, is the density on phase space
of a large ensemble of particles, $\rho$ is the spatial mass density 
induced by $f$, and $U=U(x)$ is the gravitational potential. We assume that
all the particles have the same mass, which we normalize to unity.
Solutions of this system can be viewed as equilibrium configurations
of large stellar systems such as galaxies or globular clusters,
provided that short range interactions (collisions) among the stars
are sufficiently rare to be neglected. For mathematical background
on this system we refer to \cite{Rein07}, for its astrophysical background we
refer to \cite{BT}. 
Clearly, the particle energy
\be \label{parten}
E=E(x,v) :=\frac{1}{2} |v|^2 + U(x)
\ee
satisfies the Vlasov equation (\ref{vlasov}). Hence the same is true for
any function of the form
\be \label{ansatz}
f(x,v) = \phi(E_0 - E).
\ee
Here $\phi\colon \R \to [0,\infty[$ for the moment
is at least measurable with $\phi(\eta)>0$ iff $\eta >0$ so that
the parameter $E_0<0$ is a cut-off energy: no particles whose
energy exceeds this value exist in the ensemble. With this ansatz,
\be \label{rhoUrel}
\rho(x) = \int \phi\left(E_0 - \frac{1}{2} |v|^2 - U(x)\right) dv 
=: g(E_0-U(x)),
\ee
i.e., the spatial density becomes a functional of the potential,
and the stationary Vlasov-Poisson system is reduced
to the semilinear Poisson equation
\be \label{poissonsl}
\Delta U = 4\pi g(E_0-U),\ \lim_{|x|\to \infty} U(x)=0.
\ee
{\sc Gidas, Ni}, and {\sc Nirenberg} prove in \cite{GNN} that
physically relevant solutions of the latter equation,
which in particular must lead to steady states with finite total mass,
are spherically symmetric, i.e., $U(x)=U(r)$ with $r=|x|$
and $f(x,v)=f(Ax,Av)$ for all $A\in \mathrm{SO}(3)$.
Steady states with this symmetry may in addition depend on
$L = |x\times v|^2$,
the modulus of angular momentum squared, and 
for any spherically symmetric steady state of the Vlasov-Poisson
system the particle distribution $f$ depends only on
the quantities $E$ and $L$, a fact sometimes referred to as Jeans' Theorem,
cf.\ \cite{BFH,Rein14}. We refer to the functional dependence of
$f$ on $E$ (and $L$) as the microscopic equation of state.
For the present analysis the dependence on $L$ plays no role and
is dropped.

There exist various conditions on $\phi$ which guarantee
that (\ref{poissonsl})
has solutions and that the resulting steady states
have finite mass and compact support, cf.~\cite{RammRein} and 
the references there. A necessary condition
for the latter is the existence of a cut-off energy as implemented
in the ansatz \eqref{ansatz}, cf.~\cite[Thm.~2.1]{RR00}.
It is then obvious that the spatial support of a resulting steady state
is the set where $U \leq E_0$, 
and the cut-off energy $E_0$ is the value of the potential at the boundary
of the support of the matter, provided the matter is compactly supported. 
On the other hand, we have the standard
boundary condition in (\ref{poisson}) at infinity, and due to spherical
symmetry it is natural to prescribe the value $U(0)$ of the
potential at the center. Since this is one free parameter respectively one
condition too many, we instead rewrite (\ref{poissonsl})
in terms of $y=E_0-U$ with a prescribed value at the
origin, $y(0)=\gamma>0$. Once a solution $y$ with a zero is
found we define $E_0:= \lim_{r\to \infty} y(r)$ and $U:= E_0 -y$.
In this way the cut-off energy $E_0$ is eliminated as a free
parameter and becomes part of the solution. 

Keeping an ansatz function (\ref{ansatz}) which guarantees compact support
and finite mass fixed we obtain a one-parameter family of steady states
with these properties, parameterized by $\gamma=U(R)-U(0)$ where
$[0,R]$ is the spatial support of the steady state. We refer to
$R$ as the radius of the steady state, and we define its mass by
\be \label{massdef}
M = \iint f(x,v)\, dv\, dx =  4 \pi \int_0^\infty r^2 g(y(r))\, dr.
\ee
The question we study in this paper is how $R=R(\gamma)$ and 
$M=M(\gamma)$ behave along a family of solutions parameterized by $\gamma$.

A well-known microscopic equation of state is 
the polytropic one:
\be \label{polytr}
f(x,v) = (E_0 -E)_+^k.
\ee
Here the subscript $+$ denotes the positive part, and
the resulting equation (\ref{poissonsl}) is the Lane-Emden-Fowler
equation. Steady states of finite radius and finite mass
are obtained for $-1< k < 7/2$, for $k=7/2$ the mass is still
finite but the radius is infinite, and for  $k>7/2$ the mass
is infinite, cf. \cite{BFH,Sans}.
If we fix $k$ such that mass and radius are finite, then
it is easy to see that along the corresponding one-parameter family
of polytropic steady states,
\be \label{poly_RM}
M = M(R) = C\, R^{\frac{2k-3}{2 k+1}},\ R>0,
\ee
with some positive constant $C$, provided that $k\neq -1/2$. A similar
result holds if the polytrope depends also on $L$, i.e., if the ansatz
in (\ref{polytr}) is multiplied by $L^l$. We will review the simple argument
in Appendix~\ref{polyrm}. Depending on $k$
the functional relation $M=M(R)$
is strictly increasing or strictly decreasing or constant. 

\begin{figure} 
\centering
\resizebox*{!}{0.44\textwidth}{\input{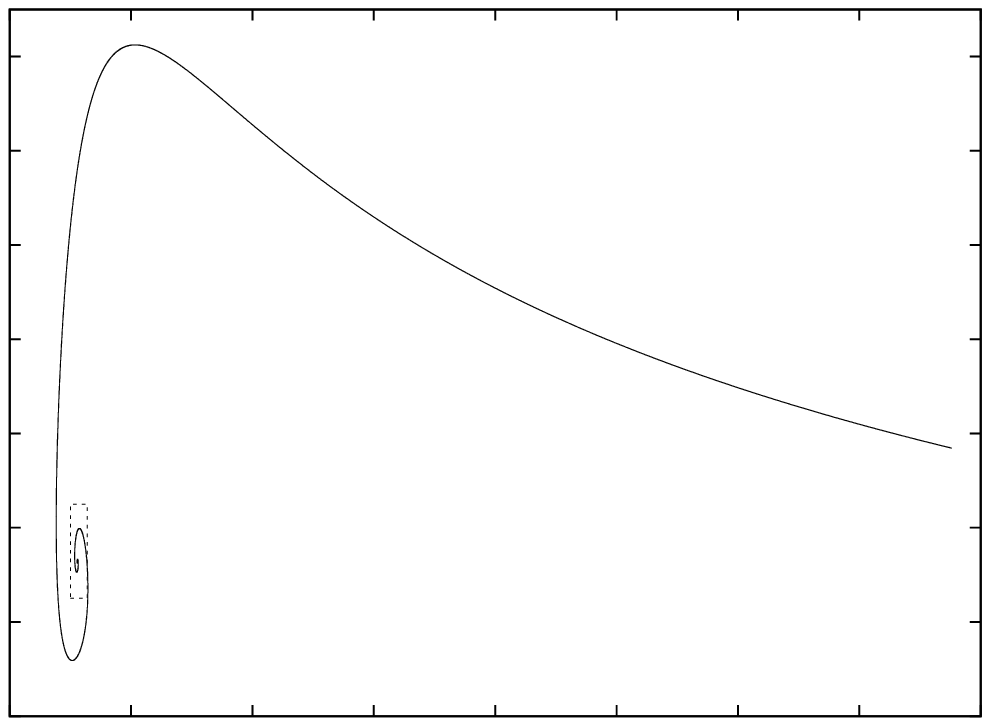}}
\hfill
{\resizebox*{!}{0.44\textwidth}{\input{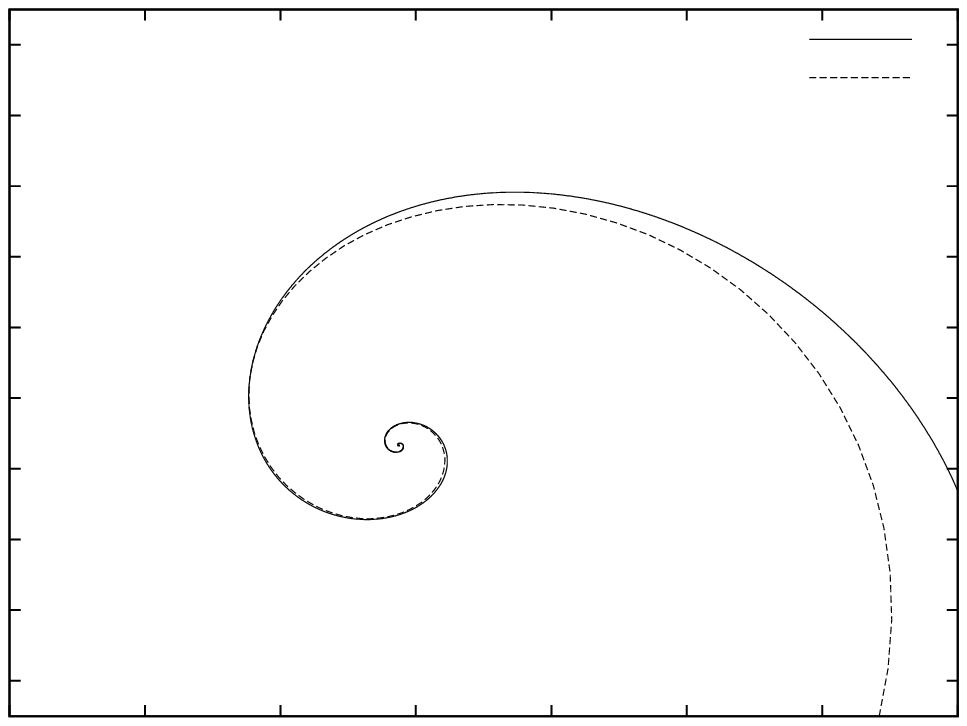}}}
\caption{Mass-radius diagram for the King model}
\label{king_fig}
\end{figure}

However, monotonicity is by no means a general feature of the 
relation between mass and radius, 
as is illustrated by the King model
\be \label{king}
f(x,v) = (e^{E_0 - E}-1)_+.
\ee
This ansatz also leads to steady states of finite mass and radius, 
since the criteria in \cite{RammRein} apply.
If one numerically computes the corresponding steady states
and their masses and radii depending on $\gamma$
and plots the results, the mass-radius diagram in Figure~\ref{king_fig}
appears: the mass-radius curve spirals into a center 
$(R_c,M_c)$ as $\gamma \to \infty$. 
The purpose of the present paper is to
give a rigorous proof of this fact for the King  and related models. 
In the blow-up of the
central part of the spiral we included the mass-radius curve predicted by our
main result, Theorem~\ref{mainthm}, with a suitable choice of parameters.

Let us put this result into perspective. 
For the Einstein-Euler system 
of general relativity {\sc Makino} shows in \cite{Mak} that  
the mass-radius diagram for families of spherically 
symmetric steady states has a spiral structure, given an
up to technical assumptions arbitrary equation of state.
It turns out that isotropic steady states of the Einstein-Vlasov system,
which is the corresponding kinetic 
model and the relativistic analogue of the Vlasov-Poisson system,
induce spherically symmetric steady states of the fluid model. 
The same correspondence holds between
isotropic steady states of the Vlasov-Poisson system and those of the 
corresponding non-relativistic fluid model, the 
Euler-Poisson system, cf.~\cite[Section~4]{Rein03}.
In \cite{AR2} this correspondence is used to show that 
for the Einstein-Vlasov system any isotropic and up to technical assumptions
arbitrary equation of state leads to a mass-radius diagram with a 
spiral structure.
But in the non-relativistic case some (microscopic)
equations of state yield simple, monotone mass-radius diagrams while
others yield a spiral structure, a fact which due to the above correspondence
is true both for the kinetic and the fluid models.  

The existence of mass-radius spirals bears on at least two broader
question for the Vlasov-Poisson system and beyond, which to a large extent
are open. Firstly, given a microscopic equation of state of the
form (\ref{ansatz}) and the corresponding one-parameter family
of steady states, how many of these steady states do have
a prescribed total mass $M$ or a prescribed radius $R$?
Does one of these parameters uniquely determine the corresponding
steady state? In the polytropic case (\ref{polytr}) any $M>0$
is attained by a unique steady state, provided $k\neq 3/2$,
cf.~(\ref{poly_RM}). For the King model (\ref{king}) the situation is very
different. There is an upper limit which $M$ can attain, below
this limit there may be several different steady states with the same
mass and different radii, and for $M=M_c$ there are infinitely
many such states. Secondly, when the steady state moves from the right
to the left of the first maximum point along the mass-radius spiral,
it is believed to change from being dynamically stable to being
unstable. This so-called Poincar\'{e} turning point principle
is far from being understood, at least in the context of kinetic
equations. For the Einstein-Vlasov system it has been
confirmed by numerical simulations, cf.~\cite{AR1}, and also 
by some partial analytic results \cite{HR1,HR2}.
However, the King steady states are non-linearly
stable in a precise sense, cf.~\cite{GR,LMR}, no matter
where they are situated along the mass-radius spiral. 
This points to the Poincar\'{e} turning point principle 
as being a challenging open problem for Vlasov type equations. 

The paper proceeds as follows. In the next section we make precise
the general framework of our analysis and formulate our main result.
This result is then proved in the following sections. 
Our analysis owes much to the paper
\cite{Mak} of {\sc Makino}. However, we give a complete,
self-contained proof. The main idea is to reformulate
the problem in such a way that the effective potential
$y=E_0-U$ becomes the independent variable and the radius $r$
and the mass within a ball of radius $r$ about the origin are
the dependent variables. This system is then put into a form which
depends on a parameter $\epsilon_\gamma$, the inverse of the central
pressure of the steady state, which is small
when $\gamma =y(0)$ is large.
The limiting system with $\epsilon_\gamma=0$ can be analyzed
using techniques for plane dynamical systems,
and the result is derived by a perturbation analysis.
The result
reported here is part of the first author's doctoral 
thesis~\cite{Ramming}.
\section{The main result}
\setcounter{equation}{0}
For our main result we require microscopic equations of
state (\ref{ansatz}), where the defining function $\phi$
is of the following form:
\be \label{phi_form}
\phi(\eta) = e^\eta - a -b\eta \ \mbox{for}\ \eta>0,\
\phi(\eta)=0\ \mbox{for}\ \eta\leq 0,\ \mbox{with}\ a,b \in \{0,1\}.
\ee
The King model (\ref{king}) is obtained for $a=1,\ b=0$,
the choice $a=b=0$ is the Woolley-Dickens model,
and $a=b=1$ is the Wilson model; our analysis
covers the choice $a=0,\ b=1$, which as far as we know
has not been considered in the astrophysics literature. This family
of admissible microscopic equations of state can be generalized,
cf.~the remark after Lemma~\ref{lemde:alpha} below.

As explained in the introduction, given such an ansatz for
$f$ the stationary Vlasov-Poisson system is reduced to
the semilinear Poisson equation (\ref{poissonsl})
with $g$ defined by (\ref{rhoUrel}), and due to spherical symmetry
the former equation can be written in terms of $y=E_0-U$ as
\be \label{ddyeq}
\frac{1}{r^2} \left(r^2 y'\right)' =
- 4 \pi g(y),
\ee
where we use that
\be \label{rhoyrel}
\rho(r) = g(y(r)),
\ee
with
\be \label{gdef}
g (y) := 2^{5/2} \pi \int_0^{y} \phi(\eta)(y-\eta)^{1/2} d\eta\
 \mbox{for}\ y>0,\
g (y) :=0 \ \mbox{for}\ y\leq 0.
\ee
It follows that $g\in C^1(\R)$, cf.~\cite{RammRein}.
Since in terms of Cartesian variables we want
potentials $U\in C^2(\R^3)$, i.e., $y\in C^2(\R^3)$,
we require that $y'(0)=0$.
Integrating (\ref{ddyeq}) once, we arrive at the equation
\be\label{yeq}
y'(r) = - \frac{m(r)}{r^2} ,
\ee
where
\be \label{mdef}
m(r) := m(r,y) = 4\pi \int_0^r s^{2} g(y(s))\, ds
\ee
is the mass within the ball of radius $r$ about the origin.
For any $\gamma >0$ the equation (\ref{yeq}) has a unique solution
$y\in C^1([0,\infty[)$ with $y(0)=\gamma$, cf.~\cite{RammRein}.
Moreover,
the results in that paper imply that in the Woolley-Dickens
and King cases (and also for the case $a=0,\ b=1$) 
this solution has a unique zero at some $R=R(\gamma)$,
which is the radius of the support of the induced steady state,
and the latter also has finite mass $M=M(\gamma)$; these assertions
hold for any $\gamma >0$. In the Wilson case these assertions
are proven in \cite{HRU} for small $\gamma$, but we are not aware of a proof
for large $\gamma$. Numerical evidence suggests
that the Wilson model has finite radius and mass also
for large $\gamma$. If this should be correct, then
the analysis below applies to the Wilson model as well;
we briefly comment on this issue at the end of Appendix~\ref{polyrm}.

In addition to the spatial mass density (\ref{rhodef})
induced by $f$ the radial pressure
\be \label{pdef}
p (x) = \int \left(\frac{x\cdot v}{r}\right)^2 f(x,v)\,dv
\ee
is important for the analysis below. If $f$ is given by an
ansatz of the form (\ref{ansatz}) a short computation shows that
\be \label{pyrel}
p(r) = h(y(r)),
\ee
where
\be \label{hdef}
h (y) := \frac{2^{7/2} \pi}{3}
\int_0^{y} \phi(\eta)(y-\eta)^{3/2} d\eta\ \mbox{for}\ y>0,\
h (y) :=0 \ \mbox{for}\ y\leq 0 ,
\ee
cf.\ \cite{RammRein,RR00}.
In order to formulate our result we define the central pressure
\be \label{pcdef}
p_c := p(0) = h(\gamma)
\ee
and introduce the additional parameter
\be \label{epsdef}
\epsilon_\gamma := \frac{1}{p_c} = \frac{1}{h(\gamma)}.
\ee
The function $h$ is strictly increasing on $[0,\infty[$
with $\lim_{y\to\infty}h(y)=\infty$, and hence
$\epsilon_\gamma \to 0$ as $\gamma \to \infty$, i.e.,
in our analysis $\epsilon_\gamma$ will play the role of a
small parameter. In addition, we define the rotation matrix
\be \label{def:Psi}
\Psi(\sigma) := \left(\begin{array}{rr}
\cos \sigma & -\sin \sigma \\
\sin \sigma & \cos \sigma \end{array}\right),\quad \sigma\in\R.
\ee
The following theorem is the main result of the present paper.
\begin{theorem} \label{mainthm}
Consider the one-parameter family of steady states induced by the
Woolley-Dickens or King ansatz respectively.
Then there exist parameters
$\bar{\gamma}>0$ and
$R_c, M_c>0$, $\tilde{\theta}\in\R^2\setminus\{0\}$, 
a function $\theta \in C\left(]\bar{\gamma},\infty[,\R^2\right)$ 
with $|\theta(\gamma)| < |\tilde \theta|/10$, and 
$\Theta\in\text{GL}(2,\R)$  such that for $\gamma > \bar{\gamma}$
the following identity holds:
\be \label{SpiralenSatz}
\left(\begin{array}{c} R \\ M \end{array}\right)(\gamma) 
= 
\left(\begin{array}{c} R_c \\ M_c \end{array}\right) + 
\epsilon_{\gamma}^{1/4}\Theta\, 
\Psi \left(\frac{\sqrt{7}}{4}\ln(\epsilon_{\gamma})\right)\cdot 
\left(\tilde{\theta} + \theta \left(\gamma\right)\right).
\ee
\end{theorem}
Some comments are in order. Firstly, to increase readability
we occasionally use $\cdot$ to indicate products of matrices
with vectors or matrices. The vector 
$\tilde{\theta} + \theta \left(\gamma\right)$ remains close to
the fixed vector $\tilde{\theta}$ as $\gamma \to \infty$, i.e., as 
$\epsilon_{\gamma} \to 0$, indeed it is possible to
replace the factor $1/10$ by any fixed factor $0<\tau<1$. Hence as
$\gamma \to \infty$ the vector given by the right hand side 
in \eqref{SpiralenSatz} spirals counterclockwise
into the center $(R_c,M_c)$. We have stated our main result for the
Woolley-Dickens and King ansatz functions, since these are
the main examples from the astrophysics literature to which
it applies. In the remark after Lemma~\ref{lemde:alpha} below we point out
possible generalizations to other ansatz functions.


%
\section{Mass and radius as dependent variables}\label{rewr_y}
\setcounter{equation}{0}
For every $\gamma>0$ there exists a unique solution
$y\in C^1([0,\infty[)$ of Eqn.~(\ref{yeq}) such that
$y(0)=\gamma$ and $y'(0)=0$. This solution is strictly decreasing,
and it has a unique zero at some radius $R=R(\gamma)$,
i.e., $y(R)=0$, $0< y(r) < \gamma$ for $0<r<R$, and $y(r)<0$ for $r>R$.
This implies that $0<\rho(r)<\rho(0)$ and $0< m(r)< m(R) = M =M(\gamma)$
for $0<r<R$. Moreover, the limit 
$\bar{y}:= \lim_{r\to \infty} y(r) \in\, ]-\infty,0[$
exists, cf.~\cite{RammRein}.
We denote the inverse function to $y\colon [0,\infty[ \to ]\bar{y}, \gamma]$
by
\[
r \colon ]\bar y, \gamma] \to [0,\infty[
\]
and define $m(y):= m(r(y))$. A short computation shows that the functions
$y\mapsto (r(y),m(y))$ satisfy the system
\be \label{ODE_(r,m)(y)}
\begin{array}{l}
\displaystyle \frac{dr}{dy}= -\frac{r^2}{m}\,,\smallskip \\
\displaystyle \frac{dm}{dy}= -4 \pi g (y)\frac{r^4}{m}\,,
\end{array}\qquad r,m>0\,,
\ee
together with the end condition
\be \label{ODE_(r,m)(y)_NB}
r(\gamma) = m(\gamma) = 0.
\ee
In terms of the solution of this system the mass and radius of the 
original steady state are now given as $M=m(0)$ and $R=r(0)$. 
For the sake of completeness we state the following result
the proof of which is straight forward.
\begin{lemma} \label{lem:(r,m)(y)=y(r)}
For every $\gamma>0$ there exists a unique, left maximal
solution $(r,m)\in C^1(]\bar{y}, \gamma[)\cap C (]\bar{y}, \gamma])$ of
\eqref{ODE_(r,m)(y)} with $\bar{y} < 0$, satisfying the end condition
\eqref{ODE_(r,m)(y)_NB}.
The function $r(y)$ has an inverse $y(r)$, 
which is the unique solution
to  \eqref{yeq} with $y(0)=\gamma$.
\end{lemma}
The key element in the proof of our main result
is to determine some $y_0>0$ such that for $\gamma > y_0$
sufficiently large the data
\[
(r_0,m_0) = (r,m)(y_0,\gamma),
\]
induced by the solutions of \eqref{ODE_(r,m)(y)}, \eqref{ODE_(r,m)(y)_NB}
exhibit the desired spiral structure as $\gamma \to \infty$. 
These data are provided by the following result,
and the spiral structure can then be shown to
persist up to $y=0$.
\begin{proposition} \label{sa:Spiralen_AW}
Let $(r,m)= (r,m)(\cdot,\gamma)$ denote the maximal solution of
\eqref{ODE_(r,m)(y)}, \eqref{ODE_(r,m)(y)_NB} which corresponds to
an ansatz of
Woolley-Dickens, King, or Wilson type respectively.
Then there exist parameters
$\bar{\gamma}>y_0>0$ and $\tilde{r}_0, \tilde{m}_0>0$, 
$\tilde{\theta}\in\R^2\setminus\{0\}$, 
a function $\theta \in C\left(]\bar{\gamma},\infty[,\R^2\right)$ 
with $|\theta(\gamma)| < |\tilde \theta|/10$, and 
$\Theta\in\text{GL}(2,\R)$  such that for $\gamma > \bar{\gamma}$
the following identity holds:
\be \label{Spiralen_AW_Vorber}
\left(\begin{array}{c} r \\ m \end{array}\right)(y_0,\gamma) 
= 
\left(\begin{array}{c} \tilde{r}_0 \\ \tilde{m}_0 \end{array}\right) + 
\epsilon_{\gamma}^{1/4}\Theta\, 
\Psi \left(\frac{\sqrt{7}}{4}\ln(\epsilon_{\gamma})\right)\cdot 
\left(\tilde{\theta} + \theta \left(\gamma\right)\right).
\ee
\end{proposition}
We first show that this result implies our main result. Its proof is then 
done in the rest of the paper. It should be noted that 
Proposition~\ref{sa:Spiralen_AW} holds also for the Wilson model;
its proof will not require that the corresponding steady states have
compact support and finite mass. In what follows, $B_r(z)$ denotes
the ball of radius $r$ centered at $z$ in $\R^n$ where $n$ will be obvious 
from the context.

\smallskip

\noindent
{\bf Proof of Thm.~\ref{mainthm}}
We restrict ourselves to the Woolley-Dickens or King case.
For the parameters $\tilde{r}_0,\tilde{m}_0, y_0>0$ 
provided by Proposition~\ref{sa:Spiralen_AW} the system
\eqref{ODE_(r,m)(y)} has a unique solution 
on the interval $[0,y_0]$ satisfying  
$(r,m)(y_0) =(\tilde{r}_0,\tilde{m}_0)$. Let  
$(r,m)(\cdot,r_0,m_0)$ denote the solution to \eqref{ODE_(r,m)(y)} 
with data $(r,m)(y_0,r_0,m_0)=(r_0,m_0)\in 
B_\delta\left((\tilde{r}_0,\tilde{m}_0)\right)$. 
For $\delta>0$ sufficiently small
this solution exists on the interval $[0,y_0]$ as well, and it is
continuously differentiable with respect to all its variables. 
We denote the derivative with respect to $(r_0,m_0)$ by
$D_{(r_0,m_0)}(r, m)$. We write
\be \label{ODE_(r,m)(y)_AW}
\left(\begin{matrix} r_0 \\ m_0 \end{matrix}\right) 
= \left(\begin{matrix} \tilde{r}_0 \\ \tilde{m}_0 \end{matrix}\right) 
+ \left(\begin{matrix} \delta_r \\ \delta_m \end{matrix}\right)
\ee
and use Taylor expansion to find that for $y\in[0,y_0]$,
\begin{eqnarray*}
\left(\begin{matrix} r \\ m \end{matrix}\right)(y,r_0,m_0) 
&=& 
\left(\begin{matrix} r \\ m \end{matrix}\right)(y,\tilde{r}_0, \tilde{m}_0)
+ 
D_{(r_0,m_0)}\left(\begin{matrix} r \\ m \end{matrix}\right)
(y,\tilde{r}_0,\tilde{m}_0) \cdot
\left(\begin{matrix} \delta_r \\ \delta_m \end{matrix}\right) \\
&&
{}+ o\left(\left|\left(\begin{matrix}\delta_r\\ \delta_m
\end{matrix}\right)\right|\right)
\end{eqnarray*}
as $\left( \delta_r, \delta_m\right)\to 0$.
Let us choose $(r_0,m_0)=(r,m)(y_0,\gamma)$ provided by
Proposition~\ref{sa:Spiralen_AW} for $\gamma >\bar{\gamma}$, which
implies that $(r,m)(\cdot,r_0,m_0)$ is the solution to \eqref{ODE_(r,m)(y)} 
which satisfies the end condition \eqref{ODE_(r,m)(y)_NB}.
In particular, if we let $y=0$ in the above Taylor expansion,
the left hand side becomes $(R,M)(\gamma)$ as desired for
Theorem~\ref{mainthm}, while the difference term
$(\delta_r,\delta_m)$ by its definition and
Proposition~\ref{sa:Spiralen_AW} takes the form
\[
\left(\begin{matrix} \delta_r \\ \delta_m \end{matrix}\right)
=
\epsilon_{\gamma}^{1/4}\Theta\, 
\Psi \left(\frac{\sqrt{7}}{4}\ln(\epsilon_{\gamma})\right)\cdot 
\left(\tilde{\theta} + \theta (\gamma)\right).
\]
We define 
$(R_c,M_c):= (r, m)(0,\tilde{r}_0, \tilde{m}_0)$ and incorporate the
$o$-term into the function $\theta$ so that the above Taylor
expansion turns into
the desired relation~\eqref{SpiralenSatz}, provided that
the matrix $D_{(r_0,m_0)} (r, m)(0,\tilde{r}_0,\tilde{m}_0)$
is regular. This follows from the
fact that $D_{(r_0,m_0)} (r, m)(\cdot,\tilde{r}_0,\tilde{m}_0)$ 
solves a linear differential equation---the variational equation
corresponding to the system \eqref{ODE_(r,m)(y)}---with
initial condition 
$D_{(r_0,m_0)} (r, m)(y_0,\tilde{r}_0,\tilde{m}_0)=\mathrm{id}$.
After redefining various parameters the proof for the fact
that Theorem~\ref{mainthm} follows from Proposition~\ref{sa:Spiralen_AW} 
is complete.
\prfe

%
\section{Proof of Proposition~\ref{sa:Spiralen_AW}}
\setcounter{equation}{0}
In this section the microscopic equation of state \eqref{ansatz}
should always be of the form \eqref{phi_form}, which includes
the Woolley-Dickens, King, and Wilson ansatz. 
We write  $\epsilon$ for the small parameter $\epsilon_{\gamma}$ defined in
\eqref{epsdef}. In the system~\eqref{ODE_(r,m)(y)}, which we need to
analyze, the independent variable is $y$. In view of
\eqref{rhoyrel} and \eqref{pyrel} we write by abuse of notation that
\be \label{Def_Spiralen_Dichte}
\rho =
\rho(y) = g (y), \quad
p =
p(y) = h(y). 
\ee 
The functions $g$ and $h$ were defined in \eqref{gdef} and
\eqref{hdef}, and they are strictly increasing
for $y\geq 0$.
In the course of the argument the quantities
\be \label{def:x_epsilon}
\frac{1}{p} =: x =: \frac{X}{p_c} = \epsilon X
\ee
are used as independent variables, where we recall from \eqref{epsdef} that
$\epsilon =1/p_c$; moving from $x$ to $X$ brings this small parameter
into the system of equations which has to be analyzed. 
By the strict monotonicity
of the functions $g$ and $h$ for positive arguments we can also
write
\[
\rho = \rho (p),
\]
and the properties of this equation of state which are
essential for the analysis below are captured in the following
auxiliary result.
\begin{lemma} \label{lemde:alpha}
Let $\rho$ and $p$ be given by \eqref{Def_Spiralen_Dichte}. Then the function
\be \label{def:alpha}
\alpha\colon ]0,\infty[\to ]0,\infty[,\quad
\alpha(\epsilon X) = \alpha(x) = 
\alpha\left(\frac{1}{p}\right) := \frac{p}{\rho(p)},
\ee
is continuously differentiable,
and $\lim_{x\to 0} \alpha(x)=1=:\alpha(0)$ so that $\alpha$ is defined as
a continuous function on $[0,\infty[$. 
Moreover, there exists an increasing function
$\omega\in C([0,\infty[)$ such that
\[
\int_0^1\omega(x)\frac{dx}{x} < \infty
\]
and
\[
\left| \alpha - 1 \right| \leq \omega .
\]
\end{lemma}
{\bf Proof.}
The function $\alpha$ is continuously differentiable, since $g$ and $h$
are and since $g$ is positive for positive arguments.
By \eqref{def:x_epsilon} and since $\lim_{y\to\infty}p(y)=\infty$,
\[
\lim_{x\to 0} \alpha(x) = 
\lim_{p\to\infty} \alpha\left(\frac{1}{p}\right)  =  
\lim_{y\to \infty} \frac{p(y)}{\rho(p(y))} .
\]
Eqns.~\eqref{gdef}, \eqref{hdef}, and \eqref{phi_form} together with 
an integration by parts imply that for $y>0$,
\begin{eqnarray*}
g(y)-h(y)
&=&
\frac{2^{5/2}\pi}{3} \int_0^y\left(e^{\eta} -a - b\eta\right)
\left(3 (y-\eta)^{1/2}- 2 (y-\eta)^{3/2}\right) d\eta \\
&=& 
\frac{2^{7/2}\pi}{3} \left[(1-a)y^{3/2} + 
\int_0^y (a + b\eta -b) (y-\eta)^{3/2}d\eta \right].
\end{eqnarray*}
Hence there exists a constant $C>0$ such that for all
$y\geq 1$,
\[
\left| \frac{p(y)}{\rho(y)} - 1 \right|  
=  \left| \frac{g(y)-h(y)}{g(y)} \right|  
\leq  \frac{Cy^{7/2}}{g (y)} =: \nu(y) .
\]
By strict monotonicity the function $p=p(y),\ y\in [0,\infty[$,
has an inverse $y=y(p),\ p\in [0,\infty[$, and we define
\[
\omega(x)=\omega\left(\frac{1}{p}\right)  :=  \nu(y(p)).
\]
By definition of $\nu$ and $\omega$, $|\alpha -1| \leq \omega$.
For $y\geq 1$ it follows that
\[
g(y)+h(y) \geq C \int_0^{y-1/2} \left(e^\eta -a- b\eta\right)\, d\eta
\geq C e^y.
\]
This implies that $\lim_{y\to\infty}\nu(y)=0$ and hence
$\lim_{x\to 0}\alpha(x)=1$. The integrability condition on $\omega$
follows from the estimate 
\[
\int_0^{1/p(1)}\omega(x)\frac{dx}{x} 
= \int_{p(1)}^{\infty}\omega\left(\frac{1}{p}\right)\frac{dp}{p} 
= C\int_1^{\infty}\frac{y^{7/2}}{h(y)}dy < \infty,
\]
and the proof is complete.
\prfe
\begin{remark}
\begin{itemize}
\item[(a)]
Lemma~\ref{lemde:alpha} captures the properties of the equation of state
which are needed for obtaining the spiral structure
in the $(R,M)$ diagram. General conditions on the ansatz function $\phi$
which guarantee these properties of $\alpha$ become rather technical.
For the ansatz function
\[
\phi(\eta):=\left\{
\begin{array}{ccl}
\eta_+^k &,& \eta \leq 1,\\
e^{\eta-1}&,& \eta >1,
\end{array}
\right.
\]
with $0<k<3/2$ Lemma~\ref{lemde:alpha} holds as well, and the results in
\cite{RammRein} show that for any $\gamma>0$ the induced steady state has 
finite mass and compact support. Hence Theorem~\ref{mainthm} holds
also for such an ansatz, and the Woolley-Dickens and King models are seen
to be members of a more general class of ansatz functions for which this 
is true.
\item[(b)]
As pointed out in the introduction,
polytropic ansatz functions yield a monotone relation
between $R$ and $M$. This corresponds to the fact that
in the polytropic case
\[
\alpha(x)= \frac{p(y)}{\rho(p(y))} =  C\frac{y^{k+5/2}}{y^{k+3/2}} 
\to \infty \ \mbox{as}\ y\to \infty
\]
as opposed to what is required in Lemma~\ref{lemde:alpha}.
\end{itemize}
\end{remark}

The proof of Proposition~\ref{sa:Spiralen_AW} proceeds in several steps.
First we observe that
by the strict monotonicity of $h$ the function $X=X(y)=p_c/p(y)$ 
has an inverse $y=y(X)$ for $1\leq X<\infty$, and $y(1) = \gamma$. 
We derive a system of differential equations for
the quantities 
\be \label{def:v(r,m)}
 v_1 = \frac{m}{r}\,,\quad v_2=4\pi r^2 p
\ee
as functions of the independent variable $X$.
The right hand side of this system depends explicitly on the
small parameter $\epsilon$, and we then analyze how solutions behave
as $\epsilon \to 0$.

\subsection{A reformulation of the system~\eqref{ODE_(r,m)(y)}}
\begin{lemma} \label{lem:ODE_v(X)}
\begin{itemize}
\item[(a)] 
Let $\alpha$ be the function defined in Lemma~\ref{lemde:alpha}, 
and let $\epsilon> 0$. The system
\begin{equation} \tag{S$_\epsilon$}
\left.\begin{array}{l}\label{ODE_(v_1,v_2)(X)}
\begin{array}{l}
\displaystyle X\frac{dv_1}{dX}  =  \frac{v_2}{v_1}-\alpha(\epsilon X), \vspace{1mm}\\
\displaystyle X\frac{dv_2}{dX}  =  \frac{v_2}{v_1}\left(2\alpha(\epsilon X)-v_1\right),
\end{array} \quad 1\leq X <\infty, \vspace{2mm}\\
\hspace*{-5.5mm}\mbox{together with the side conditions}\vspace{2mm}\\
\hspace*{-5mm}
\begin{array}{l}
\displaystyle \mbox{ (i)}\;\,
\forall \bar{X}>1 \;\exists c>0 \;\forall X\in \left[1,\bar{X}\right]:\; 
|(v_1,v_2)(X)| \leq c(X-1) \vspace{1mm} \\
\displaystyle \mbox{(ii)}\;\, 
\forall X>1:\; v_1(X),v_2(X)>0
\end{array}
\end{array}\right\}
\end{equation}
has a solution  $V=V(\cdot,\epsilon)\in C^1([1,\infty[)\cap C^2(]1,\infty[)$,
and 
\be \label{ODE_(v_1,v_2)(X)_Asympt}
\begin{array}{l}
\displaystyle V_1(X,\epsilon)  
=  2\alpha(\epsilon)\,(X-1) + O((X-1)^2), \vspace{1mm}\\
\displaystyle V_2(X,\epsilon)  
=  6\alpha^2(\epsilon)\, (X-1) + O((X-1)^2), \vspace{1mm} \\
\displaystyle \phantom{V_2(X,\epsilon)}  
=  3\alpha(\epsilon)\, V_1(X,\epsilon) + O(V_1^2(X,\epsilon)),
\end{array} \ \mbox{as}\ X\to 1.
\ee
\item[(b)] 
If $V$ solves \eqref{ODE_(v_1,v_2)(X)} for some  $\epsilon>0$,
then the function
\[
r  =  r(X)  =  \sqrt{\frac{\epsilon X V_2(X)}{4\pi}},\ X\geq 1,
\]
has an inverse $X=X(r)$ on some interval $[0,R[$ with $R\in \,]0,\infty]$. 
The function
\[
y(r)  :=  h^{-1}\left(1/(\epsilon X(r))\right),\ r\in[0,R[,
\]
solves the equation \eqref{yeq} with initial condition 
$\gamma = h^{-1}\left(1/\epsilon\right)$,
where  $m=m(r)=r V_1(X(r))$ in \eqref{yeq}.
In particular,  the solution of \eqref{ODE_(v_1,v_2)(X)} is unique. 
Moreover, $y(r)\to 0$ as $r\to R$, and hence $R$ is the radius of the
corresponding  steady state of the Vlasov-Poisson system.
\end{itemize}
\end{lemma}
{\bf Proof.}
As to part (a), we fix $\gamma >0$ and define $y$ as the maximal
solution of \eqref{yeq} with $y(0)=\gamma$ and $m$ defined as in \eqref{mdef}.
Also let $\epsilon = 1/p(\gamma)$. 
We know that $y=y(r)$ has an inverse function $r=r(y)$ and with $y=y(X)$ 
it is straight forward to check that $v=(v_1,v_2)(X)$ 
as defined in \eqref{def:v(r,m)} solves the system of differential 
equations in \eqref{ODE_(v_1,v_2)(X)}, together with the side condition~(ii). 
The asserted regularity on $]1,\infty[$ follows from the fact that
$\alpha\in C^1(]0,\infty[)$.

Continuity of $p$ and $\rho$ at $y=0$ together with the fact that
$r=r(y)\to 0$ as $y\to \gamma$ implies that
$v(X)\to 0$ as $X\to 1$. By the mean value theorem,
\[
\frac{v_2(X(r))}{v_1(X(r))} -\alpha(\epsilon X)  =  
\frac{4\pi r^3 p(r)}{m(r)} -\frac{p(r)}{\rho(r)} \to \frac{2p_c}{\rho_c}
= 2\alpha(\epsilon), \ r\to 0.
\]
Hence by the first equation in \eqref{ODE_(v_1,v_2)(X)},
\be \label{v_1_abl_asympt}
\forall \eta >0 \;\exists \delta_1>0 \;\forall X\in\ ]1, 1+\delta_1[\,:
\ X\frac{dv_1}{dX}\in B_\eta(2\alpha(\epsilon)).
\ee
By integration,
\[
v_1(X) < (2\alpha(\epsilon) + \eta)(X-1),\ X\in [1,1+\delta_1[,
\]
and the corresponding result for $v_2$ follows from \eqref{ODE_(v_1,v_2)(X)} 
for a suitable $\delta_2>0$. The functions $v_i(X)/(X-1)$ are continuous
on $]0,\infty[$ for  $i=1,2$, and the side condition (i) is established. 

We now turn to the asymptotic behavior in part (a).
From \eqref{v_1_abl_asympt} and the analogous relation for $v_2$ it follows that
\begin{eqnarray} 
v_1(X) 
&=& 
2\alpha(\epsilon)\,(X-1) + o(X-1),\ X\to 1,          \label{v_1_asympt}\\
v_2(X) 
&=& 
6\alpha^2(\epsilon)\,(X-1) + o(X-1) \nonumber \\
&=& 
3\alpha(\epsilon) \,v_1(X) + o(v_1(X)), \ X\to 1.  \label{v_2_asympt}
\end{eqnarray}
To improve these asymptotics we first note that
since $\alpha\in C^1(]0,\infty[)$ it holds that for every $\epsilon> 0$,
\begin{eqnarray}
\alpha (\epsilon X) 
&=& 
\alpha(\epsilon) + \epsilon \alpha'(\epsilon)\,(X-1) + o(X-1) \nonumber\\
&=& 
\alpha(\epsilon) + O(X-1),\ X\to 1 . \label{alpha_Taylor}
\end{eqnarray}
We define the asymptotic remainder of $v_2$ in \eqref{v_2_asympt} as
\[
\xi(X)  :=  v_2(X) - 3\alpha(\epsilon)\, v_1(X),\quad X\geq 1 .
\]
The differential equations in \eqref{ODE_(v_1,v_2)(X)} imply that
\begin{eqnarray*}
\xi(X) 
&=& 
\int_1^X \frac{1}{\sigma}
\left[ \frac{v_2(\sigma)}{v_1(\sigma)}2\alpha(\epsilon\sigma) 
- v_2(\sigma) -3\alpha(\epsilon)\frac{v_2(\sigma)}{v_1(\sigma)} 
+ 3\alpha(\epsilon)\alpha(\epsilon\sigma) \right] d\sigma    \\
&=& 
O((X-1)^2) + 
\int_1^X \left[-\alpha(\epsilon)\frac{v_2(\sigma)}{v_1(\sigma)} 
+ 3\alpha(\epsilon)^2\right] \frac{d\sigma}{\sigma}  \\
&=& 
O((X-1)^2) + 
\int_1^X \left[-\frac{\alpha(\epsilon)}{v_1(\sigma)}\right]\,
\xi(\sigma)\frac{d\sigma}{\sigma},\ X\to 1;
\end{eqnarray*}
the second equality follows from \eqref{v_2_asympt} and \eqref{alpha_Taylor}. 
By \eqref{v_1_asympt}, 
$v_1(X) \geq \frac{3}{2}\alpha(\epsilon)(X-1)$, $1\leq X\leq\bar{X}$, for some
sufficiently small $\bar{X}>1$. Hence
\begin{eqnarray*}
 |\xi(X)| 
&\leq& 
C (X-1)^2 + \frac{2}{3} \int_1^X \frac{|\xi(\sigma)|}{\sigma-1} d\sigma \\
&\leq& 
C (X-1)^2 + \frac{2}{3}\sup_{\sigma\in[1,X]} \frac{|\xi(\sigma)|}{\sigma-1} (X-1),
\ 1\leq X\leq \bar{X},
\end{eqnarray*}
and therefore
\[
\sup_{\sigma\in[1,X]}\frac{|\xi(\sigma)|}{\sigma-1} 
\leq 
C (X-1) + \frac{2}{3}\sup_{\sigma\in[1,X]} 
\frac{|\xi(\sigma)|}{\sigma-1},\ 1\leq X\leq \bar{X},
\]
which implies that
\[
\frac{|\xi(X)|}{X-1}  
\leq  
\sup_{\sigma\in[1,X]}\frac{|\xi(\sigma)|}{\sigma-1}  
\leq  
C(X-1),\ 1\leq X\leq \bar{X},
\]
and hence $\xi(X)=O((X-1)^2)$ as $X\to 1$. By definition of $\xi$,
\[
v_2(X) = 3\alpha(\epsilon)\, v_1(X) + O((X-1)^2)
= 3\alpha(\epsilon)\, v_1(X) + O(v_1^2),\ X\to 1.
\]
Together with
\[
X\frac{dv_1}{dX}(X) = \frac{v_2(X)}{v_1(X)} -\alpha(\epsilon X)  
=  2\alpha(\epsilon) + O(X-1),\ X\to 1,
\]
and after integration we find that
\[
v_1(X) = 2\alpha(\epsilon) (X-1) + O((X-1)^2),\ X\to 1,
\]
which completes the proof of the asymptotics \eqref{ODE_(v_1,v_2)(X)_Asympt}
and thus of part (a).

As to part (b), we fix some $\epsilon>0$ and consider a solution
$V$ of \eqref{ODE_(v_1,v_2)(X)} on $[1,\infty[$. The function
\[
r  =  r(X)  =  \sqrt{\frac{\epsilon X V_2(X)}{4\pi}},\ X\geq 1,
\]
is strictly increasing since
\[
\frac{dr}{dX}  =  
\frac{\epsilon\alpha(\epsilon X)}{4\pi r(X)}\frac{V_2(X)}{V_1(X)} > 0,\ X>1,
\]
and hence it has an inverse $X=X(r)$ on some interval $[0,R[$ with 
$R\in \,]0,\infty]$.
Clearly, $X(0)=1$ and $X(r)\to \infty$ as $r\to R$. If we define
$m = m(r) := r V_1(X(r))$ and  
\[
y(r)  =  h^{-1}\left(1/(\epsilon X(r))\right),\ r\in[0,R[,
\]
then the fact that $h' = g$ and Lemma~\ref{lemde:alpha} imply that
\[
\frac{dy}{dr} = 
-\frac{1}{g(y(r)) \epsilon X^2(r)}\frac{dX}{dr} 
= -\frac{4\pi r V_1(X(r))}{\epsilon X(r) V_2(X(r))}
= -\frac{m(r)}{r^2}.
\]
Moreover, the function $m(r)$ defined above satisfies the identity
\begin{eqnarray*}
\frac{dm}{dr} 
&=& 
V_1(X(r)) + r\frac{dV_1}{dX} \frac{dX}{dr} \\
&=& 
V_1(X(r)) 
+ \frac{r}{X(r)}\left[\frac{V_2(X(r))}{V_1(X(r))}-\alpha(\epsilon X(r))\right]
\frac{4\pi r}{\epsilon\alpha(\epsilon X(r))}\frac{V_1(X(r))}{V_2(X(r))} \\
&=& 
V_1(X(r)) + \frac{4\pi r^2}{\epsilon X(r) \alpha(\epsilon X(r))} 
- \frac{4\pi r^2}{\epsilon X(r) V_2(X(r))}V_1(X(r)). 
\end{eqnarray*}
In the last line the first and last term cancel by definition of $r$,
and the definition of the function $\alpha$ implies that
\[
\frac{dm}{dr} = 4\pi r^2 \rho(y(r)).
\]
Hence $m$ coincides with the function defined in \eqref{mdef},
and $y$ indeed solves \eqref{yeq} on the interval $[0,R[$ with initial condition
$\gamma  =  y(0)  =  h^{-1}\left(1/\epsilon\right)$.
Since this solution to \eqref{yeq} is unique, the proof is complete.
\prfe

\subsection{The case $\epsilon=0$}
\label{ssec:epsilon=0}

In order to analyze \eqref{ODE_(v_1,v_2)(X)} for 
$\epsilon=0$ we consider the plane, autonomous system
\be \label{S0}
\frac{dv}{ds}  = \left(
\begin{array}{c}
v_2-v_1 \\
v_2(2-v_1)
\end{array}\right) =: \zeta(v),\quad v_1,v_2\in\R.
\ee
It has the steady states
$P=(0,0)$, which is a saddle, and 
$Q=(2,2)$, which is a stable spiral. 
The set $\{v\in \R^2 \mid v_1,v_2>0\}$ is positively invariant,
and Dulac's negative criterion shows that it contains no
periodic orbits: $a(v):=1/v_2$ defines a smooth function on
this set with
\[
\mathrm{div}(a \zeta)(v) = -1/v_2 < 0 
\]
as required. Moreover,
along the line $L:= \{(v_1,8+2 v_1) \mid v_1 > 0 \}$
it holds that $ n \cdot \zeta < 0$ where $n :=(-2,1)$ is normal to $L$.
This implies that the set
\[
\mathcal M  =  \{v\in \R^2 \mid v_2 < 8+2 v_1 \} \, \cap\, ]0,12[^2
\]
is positively invariant under the flow of \eqref{S0}, and this set contains
the steady state $Q$. Since the positive $v_1$-axis
is a trajectory which is part of the stable manifold of the
saddle $P$, Poincar\'{e}-Bendixson theory implies that
$Q$ is the $\omega$-limit point of all trajectories
in $\mathcal M$.

Now we observe that the Jacobian $D\zeta (P)$ has the eigenvector
$(1,3)$ corresponding to the positive eigenvalue $2$. Hence the
unstable manifold of $P$ under the flow of \eqref{S0} contains
a trajectory which lies in the set $\mathcal M$.
Let $\tilde{V}^0=(\tilde{V}_1^0,\tilde{V}_2^0)$ denote 
a solution of \eqref{S0} which has this trajectory. 
This solution is global with 
\be \label{(i)(ii)}
\tilde{V}^0(\R)\subset \mathcal M,\ 
\tilde{V}^0(s)\to P\ \mbox{as}\ s\to -\infty,\
\mbox{and}\
\tilde{V}^0(s)\to Q\ \mbox{as}\ s\to \infty.
\ee
We define
\[
X\colon \R \to ]1,\infty[, \ 
s\mapsto \exp\left(\int_{-\infty}^s \tilde{V}_1^0(\sigma)d\sigma\right);
\]
since the solution $\tilde{V}^0$ runs in the unstable manifold of
$P$ with corresponding eigenvalue $2$, asymptotically
$\tilde{V}_1^0(s)$ behaves like $e^{2s}$ for $s\to -\infty$, and hence the 
integral in the definition of the function $X$ converges.
Moreover, $dX/ds=\tilde{V}_1^0(s)\, X>0$ so that the function 
$X=X(s)$ is strictly increasing, $X(s)\to 1$ as $s\to -\infty$, 
and $X(s)\to \infty$ as $s\to \infty$.
Let $s\colon ]1,\infty[ \,\ni X \mapsto s(X) \in \R$ denote its inverse.
Then the function $V^0(X) = \tilde{V}^0(s(X))$ is easily seen
to solve the system of differential equations in
\eqref{ODE_(v_1,v_2)(X)} for $\epsilon=0$. The unstable manifold of
the steady state $P$, in which $\tilde V^0$ runs, is tangential to the 
corresponding eigenvector $(1,3)$ so that 
$\left(V_2^0/V_1^0\right)(X)\to 3 = 3\alpha(0)$ as $X\to 1$.
By the differential equations in \eqref{ODE_(v_1,v_2)(X)}, 
$X\,dV^0/dX (X)=O(1)$ as $X\to 1$, and hence $V^0(X)=O(X-1)$ as $X\to 1$
so that $V^0$ is the solution of the problem \eqref{ODE_(v_1,v_2)(X)} 
with $\epsilon=0$, and  $V^0(X)\to Q$ as $X\to \infty$.

As in the proof of Lemma~\ref{lem:ODE_v(X)} we can derive the asymptotic
behavior
\be\label{ODE_(v^0_1,v^0_2)(X)_Asympt}
\begin{array}{l}
\displaystyle V^0_1(X) = 2(X-1) + O\left((X-1)^2\right), \vspace{1mm} \\
\displaystyle V^0_2(X) = 6(X-1) + O\left((X-1)^2\right), \vspace{1mm} \\
\displaystyle \phantom{V^0_2(X)} = 3V^0_1(X) + O\left((V^0_1)^2(X)\right),
\end{array} \quad X\to 1.
\ee

\subsection{Approximation of the center}
Our next aim is to show that as $\epsilon \to 0$ the solutions 
$V(\cdot,\epsilon)$ with $\epsilon>0$ converge to the
solution  $V^0$ of \eqref{ODE_(v_1,v_2)(X)} with $\epsilon=0$, which was
obtained in the previous section,
uniformly on every bounded interval $[1,\bar{X}]$. As a matter of fact
it would be sufficient to prove this convergence for $X$ sufficiently 
large, but due to the fact that the solutions to \eqref{ODE_(v_1,v_2)(X)}
do not satisfy regular initial conditions but the side 
condition~(i) in \eqref{ODE_(v_1,v_2)(X)}, which is singular at $X=1$,
it becomes necessary to prove a more specific estimate for $X$
close to $1$ first.
\begin{lemma} \label{LemAZ1}
Let $V(\cdot,\epsilon)$ be the solution to \eqref{ODE_(v_1,v_2)(X)}
for $\epsilon>0$ and $V^0(X)$ the solution to \eqref{ODE_(v_1,v_2)(X)} 
for $\epsilon=0$ which was constructed in Section~\ref{ssec:epsilon=0}. 
Then the following holds:
\begin{eqnarray*}
\exists\, \bar{X}>1\;
\forall\, \kappa >0 \;\exists\, \epsilon_{\kappa}>0 
&& 
\forall\, 0<\epsilon<\epsilon_{\kappa}, 
1\leq X \leq \bar{X}:\\
&&
\left|V(X,\epsilon)-V^0(X)\right|  \leq  \kappa (X-1). 
\end{eqnarray*}
\end{lemma}
{\bf Proof.}
The proof is based on the following idea.
The function $W := V(\cdot,\epsilon)-V^0$ satisfies
a certain system of differential equations. By a suitable fixpoint problem
it is shown that this system has a solution which
satisfies the estimate above, and by uniqueness this solution
must be equal to $W$. \smallskip
 
\noindent
{\em Step~1}. 
Continuity of $\alpha$ and the asymptotic expansions 
\eqref{ODE_(v_1,v_2)(X)_Asympt} and \eqref{ODE_(v^0_1,v^0_2)(X)_Asympt} 
imply that for $\epsilon>0$ sufficiently small there exists 
$\bar{X}=\bar{X}(\epsilon)>1$ such that $|W(X)|<V^0_1(X)$ for $1<X<\bar{X}$. 
Using the differential equations in \eqref{ODE_(v_1,v_2)(X)} 
for $\epsilon>0$ or $\epsilon=0$ respectively the function $W$ satisfies
on $]1,\bar{X}[$ the following system of equations;
for the sake of readability we suppress the arguments
of $V=V(X,\epsilon)$ and $V^0=V^0(X)$, we abbreviate 
$\tilde\alpha = \alpha -1$, and we use $w$ to denote the unknown in this system
while $W$ denotes the fixed difference function introduced above:
\begin{eqnarray}
X\frac{dw_1}{dX} 
&=& 
X\frac{dV_1}{dX} -X\frac{dV^0_1}{dX}  
=  \left(\frac{V_2}{V_1} -\alpha(\epsilon X)\right) 
- \left(\frac{V^0_2}{V^0_1}-1\right) \nonumber\\
&=& 
\frac{V^0_2+w_2}{V^0_1+w_1} -\frac{V^0_2}{V^0_1} 
- \tilde{\alpha}(\epsilon X) \nonumber\\
&=&
\frac{V^0_2+w_2}{V^0_1}\frac{1}{1+w_1/V^0_1} -\frac{V^0_2}{V^0_1} 
- \tilde{\alpha}(\epsilon X) \nonumber\\
&=& 
\frac{V^0_2+w_2}{V^0_1}\sum_{k=0}^\infty \left(-\frac{w_1}{V^0_1}\right)^k 
-\frac{V^0_2}{V^0_1} - \tilde{\alpha}(\epsilon X) \nonumber\\
&=&  
-\frac{V^0_2}{\left(V^0_1\right)^2}w_1  + \frac{1}{V^0_1}w_2
+ \tilde{\xi}_1(X,w,\epsilon) \label{ODE_w_1(X)},\\
X\frac{dw_2}{dX} 
&=& 
X\frac{dV_2}{dX} -X\frac{dV^0_2}{dX}  
= 
\frac{V_2}{V_1} \left(2\alpha(\epsilon X) - V_1\right) - 
\frac{V^0_2}{V^0_1} \left(2 - V^0_1\right) \nonumber\\
&=& 
2\left(\frac{V^0_2+w_2}{V^0_1+w_1} -\frac{V^0_2}{V^0_1}\right) + 
2\frac{V^0_2+w_2}{V^0_1+w_1}\tilde{\alpha}(\epsilon X) - w_2 \nonumber\\
&=& 
2\left(-\frac{V^0_2}{\left(V^0_1\right)^2}w_1  + \frac{1}{V^0_1}w_2\right)  
+ \tilde{\xi}_2(X,w,\epsilon), \label{ODE_w_2(X)}
\end{eqnarray}
where we used the fact that $|W_1| < V^0_1$ and  defined
\begin{eqnarray}
\tilde{\xi}_1(X,w,\epsilon)
&:=&
-\frac{1}{\left(V^0_1\right)^2}w_1w_2 + 
\frac{V^0_2+w_2}{V^0_1}
\sum_{k=2}^\infty \left(-\frac{w_1}{V^0_1}\right)^k - \tilde{\alpha}(\epsilon X),
\label{xi1def} \\
\tilde{\xi}_2(X,w,\epsilon)
&:=& 
2\tilde{\xi}_1(X,w,\epsilon) + 
2\left(\frac{V^0_2+w_2}{V^0_1+w_1}+1\right)\tilde{\alpha}(\epsilon X) - w_2.
\label{xi2def}
\end{eqnarray}
With
\[
\tilde{\Xi}(X)  :=  
\left(\begin{matrix}
\displaystyle - \frac{V^0_2}{\left(V^0_1\right)^2} & 
\displaystyle \ \frac{1}{V^0_1} \\
\displaystyle - \frac{2V^0_2}{\left(V^0_1\right)^2} &
\displaystyle  \ \frac{2}{V^0_1} \end{matrix}\right)(X)
\]
the system \eqref{ODE_w_1(X)}, \eqref{ODE_w_2(X)} can be written in the form
\be \label{ODE_w_xitilde}
X\frac{dw}{dX}  =  \tilde{\Xi}(X)\, w + \tilde{\xi}(X,w,\epsilon) .
\ee
Using the asymptotics \eqref{ODE_(v^0_1,v^0_2)(X)_Asympt} of $V^0$ 
we obtain the following asymptotic expansions for the
components of the matrix $\tilde{\Xi}(X)$ for $X\to 1$:
\begin{eqnarray*}
\frac{V^0_2(X)}{\left(V^0_1(X)\right)^2}  
&=&  
\frac{3 + O\left(X-1\right)}{2(X-1) + O((X-1)^2)}  
=  \frac{3}{2(X-1)}(1+O(X-1)) , \\
\frac{1}{V^0_1(X)}  
&=&  
\frac{1}{2(X-1) + O((X-1)^2)}  =  \frac{1}{2(X-1)}(1+O(X-1)) .
\end{eqnarray*}
Hence with
\[
\Xi  =  
\frac{1}{2}\left(\begin{matrix}
-3 & \ 1 \\
-6 & \ 2 \end{matrix}\right)
\]
we finally can rewrite \eqref{ODE_w_xitilde} in the form
\be \label{ODE_w_xi}
 X\frac{dw}{dX} = \frac{1}{X-1}\Xi\, w + \xi(X,w,\epsilon),
\ee
where $\xi(X,w,\epsilon) = \tilde{\xi}(X,w,\epsilon) + O(|w|)$ as $w\to 0$. 
In view of the next step we note that for every parameter value
$\epsilon>0$ the function 
$\xi(\cdot,\cdot,\epsilon)$ is defined and continuous on the set 
$\{(X,w)\in \,]1,\bar{X}[\times \R^2 \mid |w_1| < V_1^0(X)\}$.

\smallskip

\noindent
{\em Step~2}.
Due to the asymptotic expansion \eqref{ODE_(v^0_1,v^0_2)(X)_Asympt} 
we can choose $1<\bar{X}<2$ such that 
$\frac{3}{2}(X-1)\leq V_1^0(X)$ for $1\leq X\leq \bar{X}$.
For $0<\kappa<1$ we define the set of functions
\[
\mathcal{F}_{\kappa}  
=  
\left\{ w\in C\left(\left[1,\bar{X}\right]\right) \mid 
|w(X)|\leq \kappa (X-1) \right\};
\]
notice that it is sufficient to consider small $\kappa>0$ in the
assertion of the lemma.
For $w\in \mathcal{F}_{\kappa}$ the condition
$|w_1(X)| < V_1^0(X)$ holds, and 
\eqref{xi1def} and \eqref{xi2def} imply
the estimate
\be \label{Abschaetzung_xi}
|\xi(X,w(X),\epsilon)|  
\leq  
C\left(\left|\tilde{\alpha}(\epsilon X)\right| + \kappa(X-1) + \kappa^2\right),
\ 1\leq X\leq \bar{X},
\ee 
with a constant $C>0$ which does not depend on
$\epsilon$, $\bar{X}$, $\kappa$, and $w$.
Using variation of constants we turn \eqref{ODE_w_xi}
into the fixpoint problem
\[
w(X)  =  T_\epsilon(w)(X) :=
\int_1^X \Lambda(X)\Lambda(\eta)^{-1}\xi(\eta, w(\eta),\epsilon)
\frac{d\eta}{\eta},\ X \in[1,\bar{X}],
\]
where $X\frac{d\Lambda}{dX}=\frac{1}{X-1}\Xi\Lambda$. 
The matrix $\Xi$ has the eigenvalues $0$ and $-1/2$,
in particular, it can be diagonalized.
For a suitable matrix $S\in\text{GL}(2,\R)$ the matrix-valued function
\[
\Lambda(X)  =  S^{-1}\left(\begin{matrix}
 1 & 0 \\
 0 &  \sqrt{\frac{X}{X-1}} \end{matrix}\right)S
 \]
 is a fundamental matrix of the system \eqref{ODE_w_xi}, and
 \be \label{Abschaetzung_Fundamentalmatrix}
 |\Lambda(X)\Lambda^{-1}(\eta)|  \leq  C,\ 1\leq \eta \leq X \leq \bar{X},
 \ee 
 where $C>0$ again is independent of
 $\epsilon$, $\bar{X}$, $\kappa$, and $w$. 

\smallskip

\noindent
{\em Step~3.}
Let 
$\tilde{\alpha}_{\infty}(\epsilon) := 
\max_{1\leq X \leq \bar{X}}|\tilde{\alpha}(\epsilon X)|$.
Then the mapping $T_\epsilon$ satisfies for $w\in\mathcal{F}_{\kappa}$ 
the estimate
\begin{eqnarray} \label{Absch_T(w)(X)}
|T_\epsilon(w)(X)| 
&\leq& 
\int_1^X |\Lambda(X)\Lambda^{-1}(\eta)|\,
|\xi(\eta,w(\eta),\epsilon)|\frac{d\eta}{\eta} \nonumber\\
&\leq&  
C\left(\tilde{\alpha}_{\infty}(\epsilon) + \kappa(X-1) + \kappa^2\right)(X-1) 
\nonumber\\
&\leq&  
\kappa(X-1), \ 1\leq X \leq \bar{X},
\end{eqnarray}
provided $(X-1)<1/(3C)$, i.e., $\bar{X}$ is sufficiently small 
so that $1<X\leq \bar{X} < 1+1/(3C)$, 
$\kappa< \min\{1/(3C), 1\}$ is arbitrary, 
and $\epsilon_{\kappa}>0$ is such that 
$\tilde{\alpha}_{\infty}(\epsilon) < \kappa/(3C)$ for 
$0<\epsilon<\epsilon_{\kappa}$. With these choices, 
$T_\epsilon$ maps $\mathcal{F}_{\kappa}$ into itself.

We use Schauder's theorem to show that
for the above choice of parameters
the mapping $T_\epsilon \colon\mathcal F_\kappa\to \mathcal F_\kappa$ 
has a fixpoint $W^\ast$. 
The set $\mathcal{F}_{\kappa}$ is closed and convex in $C([0,\bar X])$,
so it remains to show that $T_\epsilon$ is compact.
We first show that  $T_\epsilon$ is continuous.
To this end, let  $(w_n)_{n\in\N}\subset \mathcal F_\kappa$ 
be a sequence which converges uniformly to some 
$w\in \mathcal F_\kappa$, and let $\delta>0$.
There exist constants $C_1, C_2>0$, which do not depend on $X$,
such that for $1\leq X\leq\bar{X}$,
\begin{eqnarray*}
|T_\epsilon(w_n)
-
T_\epsilon(w)|(X) 
&=& 
\int_{1}^{X} 
\left|\Lambda(X)\Lambda^{-1}(\eta)
\left[\xi(\eta,w_n(\eta),\epsilon)-\xi(\eta,w(\eta),\epsilon)\right]\right|
\frac{d\eta}{\eta}\\
&\leq&  
\ C\int_1^{\bar{X}}
\left|\xi(\eta,w_n(\eta),\epsilon)-\xi(\eta,w(\eta),\epsilon)\right|
\frac{d\eta}{\eta} \\
&\leq&  
C_1\left(\tilde{X} -1\right)\\ 
&&
{}+ 
C_2\int_{\tilde{X}}^{\bar{X}}
\left|\xi(\eta,w_n(\eta),\epsilon)-\xi(\eta,w(\eta),\epsilon)\right|d\eta \\
&<&
\frac{\delta}{2} + \frac{\delta}{2}  =  \delta,\ n>N,
\end{eqnarray*}
provided that
\[
1\leq\tilde{X}<\min\left\{1+\frac{\delta}{2C_1}, \bar{X}\right\}
\]
and $N\in\N$ is sufficiently large so that
\[
\max_{\eta\in [\tilde{X},\bar{X}]}  
\left|\xi(\eta,w_n(\eta),\epsilon)-\xi(\eta,w(\eta),\epsilon)\right| 
< 
\frac{\delta}{2C_2(\bar{X}-1)},\ n>N.
\]
This proves that $T_\epsilon$ is continuous.

It remains to show that $T_\epsilon$ maps bounded sets
into relatively compact ones. Since $\mathcal F_\kappa$ itself is
uniformly bounded we only need to check for equicontinuity. Thus let
$w\in\mathcal F_\kappa$ and
$1< X_1\leq X_2\leq \bar{X}$. Then
\begin{eqnarray*}
|T_\epsilon(w)(X_2)
&-&
T_\epsilon(w)(X_1)| \\
&&
\leq  \left|\int_{X_1}^{X_2} \Lambda(X_2)\Lambda^{-1}(\eta)
\xi(\eta,w(\eta),\epsilon)\frac{d\eta}{\eta}\right|\\
&&
\qquad {}+ 
\left|\int_1^{X_1} 
\left[\Lambda(X_2)\Lambda^{-1}(\eta)-\Lambda(X_1)\Lambda^{-1}(\eta)\right]
\xi(\eta,w(\eta),\epsilon)\frac{d\eta}{\eta}\right|\\
&&
\leq  C\int_{X_1}^{X_2}d\eta 
+ C\int_1^{X_1} 
\left|\Lambda(X_2)\Lambda^{-1}(\eta)-\Lambda(X_1)\Lambda^{-1}(\eta)\right|
d\eta \\
&&
\leq  C(X_2 -X_1)  + 
C \left|\sqrt{\frac{X_2}{X_2-1}} - \sqrt{\frac{X_1}{X_1-1}}\right|
\int_1^{X_1}\!\sqrt{\eta-1}\,d\eta \\
&&
\leq C(X_2 -X_1),
\end{eqnarray*}
where the constant $C>0$ is independent of  $X_1$, $X_2$, and $w$.
Hence the set $T_\epsilon(\mathcal F_\kappa)$ is equicontinuous,
and by Arzel\`{a}-Ascoli, it is relatively compact.

By Schauder's theorem  the mapping
$T_\epsilon \colon \mathcal F_\kappa\to \mathcal F_\kappa$
has a fixpoint $W^\ast$. This function solves \eqref{ODE_w_xi}, and since
solutions to \eqref{ODE_(v_1,v_2)(X)} are unique,
$V(\cdot,\epsilon) - V^0 = W = W^\ast$. The latter function
by construction lies in the set $\mathcal F_\kappa$, which 
implies the estimate claimed in the lemma, and the proof is complete. 
\prfe

We need to get rid of the restriction to sufficiently small $\bar{X} >1$.
Let $\bar{X}$ be as obtained in Lemma~\ref{LemAZ1},
and let $\bar{\bar{X}}>\bar{X}$ be arbitrary. 
According to Section~\ref{ssec:epsilon=0},
$V^0(X)\to Q$ as $X\to\infty$.
Hence
\[
d := \text{dist}\left(V^0([\bar{X},\infty[),\, 
\R\times\{0\} \cup \{0\}\times\R\right) > 0,
\]
so there exists a compact set $K\subset ]0,\infty[^2$ such that
\[
\text{dist}\left(V^0([\bar{X},\infty[), \partial K\right)
>\frac{d}{2}.
\]
On the set
$\{(X,v,\epsilon)\in \R^4 \mid \bar{X}\leq X\leq \bar{\bar{X}},\
v\in K,\ 0\leq \epsilon\leq 1\}$ 
the right hand side $\zeta_\epsilon$ of the system in
\eqref{ODE_(v_1,v_2)(X)}
is Lipschitz continuous with respect to $v$, and
\[
|\zeta_\epsilon (X,v) - \zeta_0(X,v)| \to 0,\ \epsilon\to 0,
\]
uniformly in $\bar{X}\leq X\leq \bar{\bar{X}}$ and $v\in K$. 
Continuous dependence on parameters implies the following result.
\begin{lemma}
Let $V(\cdot,\epsilon)$ and $V^0$ be as in Lemma~\ref{LemAZ1}, and let
$\bar{X}>1$. Then 
$V(X,\epsilon) \to V^0(X)$ as $\epsilon\to 0$, uniformly 
in $X\in [1,\bar X]$.
\end{lemma}

\subsection{A linearization lemma}
\label{sec:RedLem}

The final step in the analysis consists in understanding the
behavior of the solutions to \eqref{ODE_(v_1,v_2)(X)} as $\epsilon \to 0$. 
The key to this is the precise relation between solutions to that equation
and those of the initial value problem
for a corresponding linear system of the form
\be \label{ivpsinglin}
t \frac{du}{dt} = B u,\ u(t_0)=u_0.
\ee
where $B \in \R^{2\times 2}$, $t_0>0$, and $u_0\in \R^2$.
We use $t$ and $u$ for the independent and dependent variable
here, in order to make this section notationally independent
of the rest of the paper and possibly useful also in other situations.
First we collect some obvious facts concerning \eqref{ivpsinglin}.
For $t>0$ we define 
\[
t^B := \exp(\ln(t) B) .
\]
Then the function
\[
]0,\infty[\, \ni t\mapsto \left(t/t_0\right)^B u_0
\]
is the unique, maximal
solution of the initial value problem \eqref{ivpsinglin}.
Assume now that  $B$ has eigenvalues $\mu \pm i\nu$ with $\nu >0$
so that there exists a regular matrix $S \in\text{GL}(2,\R)$ such that
\[
S B S^{-1} = 
\left(\begin{matrix}
\mu & -\nu \\
\nu & \mu \end{matrix}\right) = \mu E + \left(\begin{matrix}
0 & -\nu \\
\nu & 0 \end{matrix}\right).
\]
Then
\be \label{tBjordan}
t^B = t^\mu S \Psi(\nu\ln t) S^{-1}
\ee
with $\Psi$ defined as in \eqref{def:Psi}. There exists a constant
$C=C_B>0$ such that for all $t>0$,
\be \label{tBest}
\left|t^B\right|\leq C_B t^\mu,\ \left|t^{-B}\right| \leq C_B t^{-\mu};
\ee
notice that $-B$ has the eigenvalues $-\mu \pm i\nu$.
We now state our linearization lemma. 
\begin{theorem}  \label{RedLem}
Consider the two dimensional system of ordinary differential equations
\be \label{RedLem_u(t)}
t\frac{du}{dt} = a_0(t) - A(t)\, u + a_2(t, u)
\ee
where $a_0\in C([0,T];\,\R^2)$, $A\in C([0,T];\,\R^{2\times 2})$, 
$a_2\in C([0,T]\times B_\delta(0);\R^2)$, $a_2$ is twice continuously
differentiable with respect to $u$, and $T, \delta >0$. 
Moreover, let $A(0)$ be regular with eigenvalues $\mu\pm i\nu$, $\mu,\nu>0$,
and let $a_0$, $A$, and $a_2$ satisfy for all $t\in [0,T]$
and $u\in B_\delta(0)$ the conditions
\begin{align*}
& |a_0(t)| 
\leq 
\omega(t) ,      \tag{C0} \label{RedLem_C0} \\
& |A(t) - A(0)| 
\leq 
\omega(t) , \tag{C1} \label{RedLem_C1} \\
& |a_2(t,u)| 
\leq
C|u|^2 ,        \tag{C2} \label{RedLem_C2} 
\end{align*}
with some constant $C>0$ and some
increasing function $\omega\in C([0,T])$ which satisfies the integrability 
condition
\[
\int_0^T\omega(\sigma)\frac{d\sigma}{\sigma} < \infty .
\]
Then there exist 
$0<\bar{t}<T$, $0<\tilde{\delta}<\delta$, and functions
$\gamma_0\in C([0,\bar{t}\,]; \R^2)\cap C^1(]0,\bar{t}\,]; \R^2)$, 
$\Gamma\in C([0,\bar{t}\,]; \R^{2\times 2})\cap 
C^1(]0,\bar{t}\,];\R^{2\times 2})$, 
and $\gamma_2 \in C^1(\Omega;\R^2)$ where
\[
\Omega := \{(t, t_0, z_0) \in \,]0,\bar t\,]^2\times B_{\tilde{\delta}}(0) 
\mid  t\geq t_0\},
\]
such that the following is true:
For every choice of data $t_0\in \,]0,\bar{t}[$ and
$z_0 \in B_{\tilde{\delta}}(0)$ and denoting 
$z(t):=\left(t/t_0\right)^{-A(0)} z_0$,
the function
\[
u(t) = \gamma_0(t) + \Gamma(t) \left(z(t) + \gamma_2(t,t_0, z_0)\right),\
t\in [t_0, \bar{t}\,], 
\]
is a solution of \eqref{RedLem_u(t)}. The functions 
$\gamma_0$ and $\Gamma$ satisfy for $0\leq t \leq \bar{t}$ the estimates
\beas
|\gamma_0(t)| 
&\leq& 
C t^{-\mu} \int_0^t \sigma^\mu \omega(\sigma)\frac{d\sigma}{\sigma},\\
|\Gamma(t) - E|
&\leq& 
C\int_0^t \omega(\sigma)\frac{d\sigma}{\sigma} 
\eeas
with some constant $C>0$.
The function $\gamma_2$ has the properties that
\[
|\gamma_2(t,t_0,z_0)| \leq C|z_0|^2 \left(t/t_0\right)^{-\mu}, 
\ (t, t_0, z_0)\in \Omega
\]
with a constant $C>0$ which does not depend on 
$t_0$ and $z_0$, and 
\[
\gamma_2(t_0,t_0,z_0) = 0,\ 
(t_0, t_0, z_0)\in \Omega. 
\]
\end{theorem}
Clearly, $\gamma_0(t)\to 0$ and $\Gamma(t)\to E$ as $t\to 0$,
which is what we actually use in the rest of the paper,
but requiring these weaker conditions does not seem to
allow for a simpler proof of Theorem~\ref{RedLem}.
The latter is an adaptation of \cite[Thm.~2]{Mak} 
to the present situation. The proof in \cite{Mak} relies on power 
series expansions and hence requires high regularity of the given
functions. Our result makes no such demands. Its proof is 
postponed to Appendix~A in order not to interrupt the line of argument
towards our main result.

\subsection{Concluding the proof of Prop.~\ref{sa:Spiralen_AW}}
 
We already know that on the one hand,
$V^0(X)$ converges to the stable spiral point $Q=(2,2)$ as $X\to \infty$.
On the other hand, on any given compact interval $[1,\bar{X}]$
the difference $V(X,\epsilon)-V^0(X)$ 
becomes as small as we wish when $\epsilon$ is taken sufficiently small. 
Hence $V(X,\epsilon)$ is close to $Q$ for $X$ large and 
$\epsilon$ small. To make this precise,
the main step will be  to use Theorem~\ref{sec:RedLem} 
in order to reduce the system obeyed by
$V=V(x/\epsilon,\epsilon)$ for $\epsilon>0$
to a simpler system from which the spiral behavior can then be deduced.
\begin{lemma} \label{lem:AW_v(x/epsilon,epsilon)}
Let $0<\tau<1$, $\Psi$ as defined \eqref{def:Psi}, and let $V(X,\epsilon)$
denote the unique solution to \eqref{ODE_(v_1,v_2)(X)} with $\epsilon>0$.
Then there exist parameters
$0<\bar{\epsilon}<\bar{x}$, $b_0\in \,]0,\infty[^2$,
$b_2\in\R^2\setminus\{0\}$, 
$b_\epsilon\in B_{\tau|b_2|}(0)$ depending continuously on $\epsilon$,
and $B\in\text{GL}(2,\R)$ such that for $0<\epsilon<\bar{\epsilon}$ 
the following identity holds:
\be \label{lem:AW_v(x/epsilon,epsilon)_V}
V\left(\frac{\bar{x}}{\epsilon},\epsilon\right) = 
b_0 + \epsilon^{1/4} B \Psi\left(\frac{\sqrt{7}}{4}\ln\epsilon\right)\cdot
\left(b_2+b_\epsilon\right).
\ee 
\end{lemma}
{\bf Proof.}
As in the proof of Lemma~\ref{LemAZ1} we consider for $\epsilon>0$
the difference $W(x) = V(x/\epsilon, \epsilon)-Q$ such
that $|W_1|<2$ on some interval $I\subset ]0,\infty[$. For $x\in I$ 
the function $W$ satisfies the following system of differential equations,
where we recall the abbreviation $\tilde\alpha:=\alpha -1$:
\begin{eqnarray}
x\frac{dw_1}{dx} 
&=& 
\frac{2+w_2}{2+w_1} -\alpha(x) 
=
\frac{2+w_2}{2}\sum_{k=0}^\infty \left(-\frac{w_1}{2}\right)^k - 
\alpha(x) \nonumber\\
&=& 
- \tilde{\alpha}(x) - \frac{1}{2}\left(w_1-w_2\right)
-\frac{1}{4}w_1w_2 + 
\frac{2+w_2}{2}\sum_{k=2}^\infty \left(-\frac{w_1}{2}\right)^k,\ 
\label{AW_w_1(x)}\\
x\frac{dw_2}{dx} 
&=& 
\frac{2+w_2}{2+w_1} \left(2\alpha(x) - 2 - w_1\right) \nonumber\\
&=& 
2\tilde{\alpha}(x) + \left(\tilde{\alpha}(x)w_2 - \alpha(x)w_1\right) 
\nonumber\\
&&
{}-\frac{1}{2}\alpha(x)w_1w_2 + 
(2+w_2)\alpha(x)\sum_{k=2}^\infty \left(-\frac{w_1}{2}\right)^k .
\label{AW_w_2(x)}
\end{eqnarray}
On the other hand, if $w$ is a solution of this system
on some interval $]x_1,x_2[\,\subset ]0,\infty[$,
then $V(X,\epsilon) = Q + w(\epsilon X)$ defines for $\epsilon>0$
a solution to the system of differential equations in
\eqref{ODE_(v_1,v_2)(X)} on $]x_1/\epsilon,x_2/\epsilon[$, 
which however need not
satisfy the side conditions in \eqref{ODE_(v_1,v_2)(X)}.
But if for an $\epsilon>0$ this function $V$ 
coincides at some point with the solution of 
\eqref{ODE_(v_1,v_2)(X)}, 
then by uniqueness for \eqref{ODE_(v_1,v_2)(X)} this holds
on all of $]x_1/\epsilon,x_2/\epsilon[$. In what follows, we
aim to identify solutions $w$ of the above system which fit
in this sense.

To this end we define 
$a_0\in C([0,\infty[;\R^2)$, 
$A\in C([0,\infty[;\R^{2\times 2})$, 
and $a_2\in C([0,\infty[\times B_1(0); \R^2)$ by
\be \label{AW_Def_a0_A}
a_0(x)  
=  \left(\begin{matrix}
- \tilde{\alpha}(x) \\
2\tilde{\alpha}(x) \end{matrix}\right)\,,\quad
A(x)  =  \frac{1}{2}\left(\begin{matrix}
1 & -1 \\
2\alpha(x) &\ -2\tilde{\alpha}(x) \end{matrix}\right),
\ee  
and
\be \label{AW_Def_a2}
a_2(x,w)  =  
\left(\begin{matrix}
\displaystyle -\frac{1}{4}w_1w_2 + 
\frac{2+w_2}{2}\sum_{k=2}^\infty \left(-w_1/2\right)^k \\
\displaystyle - \frac{1}{2}\alpha(x)w_1w_2 + 
(2+w_2)\alpha(x)\sum_{k=2}^\infty \left(-w_1/2\right)^k 
\end{matrix}\right),
\ee 
and the system \eqref{AW_w_1(x)}, \eqref{AW_w_2(x)} can be written as
\be \label{ODE_AW_w(x)}
x\frac{dw}{dx}  =  a_0(x) - A(x)\, w + a_2(x, w).
\ee 
We wish to apply Theorem~\ref{RedLem} to this system.
Using \eqref{AW_Def_a0_A}, \eqref{AW_Def_a2}, and Lemma~\ref{lemde:alpha} 
one can verify that the functions $a_0$, $A$, and $a_2$ 
satisfy the conditions \eqref{RedLem_C0}-\eqref{RedLem_C2} 
in Theorem~\ref{RedLem}. Moreover, the matrix
\[
A(0)  =  \frac{1}{2}
\left(\begin{matrix}
1 & -1 \\
2 & 0 \end{matrix}\right)
\]
has the eigenvalues
\[
\lambda_{1/2}  =  \frac{1}{4} \pm \frac{\sqrt{7}}{4}i .
\]
Hence there exists a regular matrix $T\in \R^{2 \times 2}$ such that
\be \label{AW_A_Jordantrafo}
T A(0) T^{-1}  =  \frac{1}{4}
\left(\begin{matrix}
1 & -\sqrt{7} \\
\sqrt{7} & 1 \end{matrix}\right).
\ee 
Clearly, there exist constants $\sigma_1, \sigma_2>0$ such that
\be \label{AW_Tsigma}
\sigma_1 |v| \leq  |Tv|  \leq  \sigma_2 |v|,\ v\in\R^2.
\ee 
According to Theorem~\ref{RedLem} there exist parameters 
$\bar{x}>0$ and $0<\tilde{\delta}<1$, and functions 
$\gamma_0$, $\Gamma$, and $\gamma_2$ such that for every
choice of data
$x_0\in \,]0,\bar{x}\,[$ and $z_0 \in B_{\tilde{\delta}}(0)$
and with
$z(x):=\left(x/x_0\right)^{-A(0)} z_0$, which is the solution to
\be \label{RedLinSys}
x\frac{dz}{dx}  =  -A(0)\, z,\ z(x_0)=z_0,
\ee 
the definition
\be \label{TrafoMakino}
w(x)  :=  \gamma_0(x) + \Gamma(x)\, \big(z(x) + \gamma_2(x,x_0, z_0)\big),
x\in [x_0, \bar{x}\,],
\ee 
yields a solution to \eqref{ODE_AW_w(x)}. Moreover,
the function $\gamma_2$ satisfies the estimate
\be \label{AW_gamma2_Absch}
|\gamma_2(x,x_0, z_0)|  \leq  C |z_0|^2 \left(x/x_0\right)^{-1/4},
\ x\in[x_0,\bar{x}],
\ee 
with a constant $C>0$ which is independent of the data
$x_0$ and $z_0$.

The choice $z_0=0$ in \eqref{TrafoMakino} shows that
$w=\gamma_0$  solves \eqref{ODE_AW_w(x)}.
This implies that for every $\epsilon\in \,]0,\bar x[$ the function 
\[
V^s(X,\epsilon) := Q + \gamma_0(\epsilon X),\ X\in [1,\bar{x}/\epsilon],
\]
solves the system of differential equations in 
\eqref{ODE_(v_1,v_2)(X)}, and for fixed  $X$,
$V^s(X,\epsilon)\to Q$ as $\epsilon\to 0$, 
since $\gamma_0(x)\to 0$ as $x\to 0$. In particular,
$V^s(X,\epsilon) \in \,]0,\infty[^2$ as required for solutions of
\eqref{ODE_(v_1,v_2)(X)}, provided  that $\bar x$ is sufficiently small.

It remains to find,
for solutions $V(X,\epsilon)$ of \eqref{ODE_(v_1,v_2)(X)}
with $\epsilon>0$ which are sufficiently close to $V^s(X,\epsilon)$,  
suitable data for the initial value problem \eqref{RedLinSys}.
For $\sigma_1$ and $\sigma_2$ as in \eqref{AW_Tsigma}, $0<\tau<1$,
$C>0$ as in \eqref{AW_gamma2_Absch} and $C_{A(0)}>0$ according to
\eqref{tBest} we define
\[
\delta  :=  
\min\left\{\tilde{\delta}, \frac{\sigma_1\tau}{2\sigma_2 C C_{A(0)}}\right\}.
\]
By Theorem~\eqref{RedLem}, $\Gamma(x)\to E$ as $x\to 0$. 
Hence we can take $\bar{x}>0$ sufficiently small so that
the inverse matrix $\Gamma^{-1}(x)$ exists and satisfies the 
estimate $|\Gamma^{-1}(x)|<2$ for $0\leq x\leq \bar{x}$. 
Next let $X^*>1$ be sufficiently large so that 
\[
|V^0(X^*)-Q|<\frac{\delta}{6},
\] 
and let $0<\bar{\epsilon}\bar x$ be sufficiently small so that
$\bar{\epsilon} X^* < \bar{x}$ and
\[
|V(X^*,\epsilon) - V^0(X^*)|,\ | Q - V^s(X^*,\epsilon)|<\frac{\delta}{6},\ 
0<\epsilon<\bar{\epsilon} .
\]
Then
\[
|V(X^*,\epsilon) - V^s(X^*,\epsilon)| 
<
\frac{\delta}{2}, \ 0<\epsilon<\bar{\epsilon}.
\]
For $0<\epsilon<\bar{\epsilon}$ we define
\be \label{AW_AW}
x_0^\epsilon  :=  \epsilon X^*,\ 
z_0^\epsilon  
:=  \Gamma^{-1}(\epsilon X^*)\left(V(X^*,\epsilon)-V^s(X^*,\epsilon)\right),
\ee 
as the desired data for the initial value problem
\eqref{RedLinSys}; clearly, $z_0^\epsilon\in B_{\tilde{\delta}}(0)$). 
Moreover
\[
z_0^\epsilon\to \tilde{z}_0 = V^0(X^\ast) - Q \neq 0\  \mbox{as}\ 
\epsilon\to 0. 
\]
For $0<\epsilon<\bar{\epsilon}$ the function
\[
w(x,\epsilon)  =  \gamma_0(x) + 
\Gamma(x) \left(\left(\frac{x}{x_0^\epsilon}\right)^{-A(0)}
z_0^\epsilon + \gamma_2(x,x_0^\epsilon, z_0^\epsilon)\right)
\]
solves \eqref{ODE_AW_w(x)} on the interval $[\epsilon X^*, \bar x]$,
and $V(X,\epsilon) = Q+ w(\epsilon X, \epsilon)$ coincides on 
$[X^*, \bar{x}/\epsilon]$ with the solution to \eqref{ODE_(v_1,v_2)(X)} 
for $\epsilon>0$. In particular this implies that for $x=\bar{x}$ and 
$0<\epsilon<\bar{\epsilon}$,
\be \label{AW_V(barx/epsilon)}
 V\left(\frac{\bar{x}}{\epsilon},\epsilon\right)  
=  Q + \gamma_0(\bar x) 
+ \Gamma(\bar{x})
\left(\left(\frac{\bar{x}}{\epsilon X^*}\right)^{-A(0)} z_0^\epsilon + 
\gamma_2(\bar{x},x_0^\epsilon, z_0^\epsilon)\right).
\ee 
With $\mu=1/4$ and $\nu=\sqrt{7}/4$ Eqn.~\eqref{tBjordan}
implies that
\[
\epsilon^{A(0)} 
= 
\epsilon^\mu T^{-1} \Psi (\nu \ln\epsilon ) T.
\]
We define $b_0:= Q + \gamma_0(\bar x)$, $B:=  \Gamma(\bar{x}) T^{-1}$, 
$b_2:= \left(\bar{x}/ X^*\right)^{-A(0)} \tilde z_0\in \R^2\setminus\{0\}$,
and
\be \label{beps}
b_\epsilon:= \left(\bar{x}/ X^*\right)^{-A(0)} 
\left(z_0^\epsilon-\tilde z_0\right)
+ \epsilon^{-\mu} \Psi(-\nu\ln\epsilon)
T\,\gamma_2(\bar{x},x_0^\epsilon, z_0^\epsilon).
\ee
Then \eqref{AW_V(barx/epsilon)}
turns into the asserted relation \eqref{lem:AW_v(x/epsilon,epsilon)_V},
and it remains to estimate $b_\epsilon$ against $\tau |b_2|$.
The first term in \eqref{beps} can be estimated against
$\tau |b_2|/3$ for $\epsilon$ sufficiently small, since
$z_0^\epsilon \to \tilde z_0$ as $\epsilon\to 0$. 
By \eqref{tBest}, \eqref{AW_Tsigma}, and \eqref{AW_gamma2_Absch},
\begin{eqnarray*}
\Bigl|\epsilon^{-\mu} \Psi(-\nu\ln\epsilon) T
&&
\gamma_2(\bar{x},x_0^\epsilon, z_0^\epsilon)\Bigr|\\
&&
\leq 
\epsilon^{-\mu}\sigma_2 C \left(\frac{\bar{x}}{X^*\epsilon}\right)^{-\mu}
\left|z_0^\epsilon\right|^2
\leq  
C \sigma_2 \delta  \left(\frac{\bar{x}}{X^*}\right)^{-\mu}
\left|z_0^\epsilon\right|\\
&&
\leq  
\frac{\tau}{2}\sigma_1 \left(\frac{\bar{x}}{X^*}\right)^{-\mu}
\left|z_0^\epsilon\right|
\leq
\frac{\tau}{2}\left|T\left(\frac{\bar{x}}{X^*}\right)^{-A(0)}
z_0^\epsilon\right|.
\end{eqnarray*}
Hence due to the convergence of $z_0^\epsilon$ the second term in \eqref{beps}
can be estimated against $2\tau |b_2|/3$ for $\epsilon$ sufficiently small,
and the proof is complete.
\prfe

\noindent
For the proof of Proposition~\ref{sa:Spiralen_AW} we need to transform
the dependent variable $v$ back into $(r,m)$, which are the dependent
variables of interest for our result, i.e., for fixed $p$ we need to
solve \eqref{def:v(r,m)}  for $(r,m)$.

\noindent
{\bf Proof of Prop.~\ref{sa:Spiralen_AW}.}
Lemmas~\ref{lem:(r,m)(y)=y(r)} and \ref{lem:ODE_v(X)} show that
for $\gamma>0$ and
$\epsilon_\gamma = 1/h(\gamma)$
the system \eqref{ODE_(r,m)(y)} with the side condition \eqref{ODE_(r,m)(y)_NB}
is equivalent to \eqref{ODE_(v_1,v_2)(X)} for $\epsilon=\epsilon_\gamma$. 
For $\bar{\epsilon}>0$ as provided by 
Lemma~\ref{lem:AW_v(x/epsilon,epsilon)}
we restrict ourselves to $0<\epsilon_\gamma<\bar{\epsilon}$, i.e., to
\[
\gamma >\bar{\gamma} = h^{-1}\left(1/\bar{\epsilon}\right).
\]
We choose $p=1/x=1/\bar{x}$ in the system \eqref{def:v(r,m)} and solve 
it for $(r,m)$ so that 
\[
r = r(\bar x,v) := \sqrt{\frac{\bar{x}v_2}{4\pi}},\ 
m = m(\bar x,v) := \sqrt{\frac{\bar{x}v_2}{4\pi}}v_1.
\]
For an arbitrary, but fixed point $\bar{v}\in \,]0,\infty[^2$ 
Taylor expansion implies that
\begin{eqnarray*}
\left(\begin{matrix} r \\ 
m \end{matrix}\right)(\bar{x},v) 
&=& 
\left(\begin{matrix} r \\ 
m \end{matrix}\right)(\bar{x},\bar{v}) + D_v\left(\begin{matrix} r \\ 
m \end{matrix}\right)\Bigr|_{x=\bar{x}, v=\bar{v}}\, (v-\bar{v}) 
+ o(|v-\bar{v}|) \\
&=& 
\left(\begin{matrix} r \\ 
m \end{matrix}\right)(\bar{x},\bar{v}) +
 \sqrt{\frac{\bar{x}}{4\pi}}
\left(\begin{matrix} 0 & \frac{1}{2\sqrt{\bar{v}_2}} \\ 
\sqrt{\bar{v}_2} & \frac{\bar{v}_1}{2\sqrt{\bar{v}_2}} 
\end{matrix}\right)\, (v-\bar{v}) + o(|v-\bar{v}|) \\
&=& 
\left(\begin{matrix} \bar{r} \\ 
\bar{m} \end{matrix}\right) + J\, (v-\bar{v}) + o(|v-\bar{v}|),
\ v\to \bar{v},
\end{eqnarray*}
with the obvious definitions
for $\bar{r}$, $\bar{m}>0$, and $J\in \text{GL}(2,\R)$.
If we substitute $\bar{v}=b_0$ and $v=V(\bar x/\epsilon,\epsilon)$
as provided by Lemma~\ref{lem:AW_v(x/epsilon,epsilon)}, then
since $V(\cdot,\epsilon)$ solves \eqref{ODE_(v_1,v_2)(X)},
$(r,m)(\bar{x},v)=(r,m)(y_0,\gamma)$ is a solution of
\eqref{ODE_(r,m)(y)}, \eqref{ODE_(r,m)(y)_NB}, where 
$0<y_0 = h^{-1}(1/\bar x) <  h^{-1}(1/\bar \epsilon) = \bar\gamma$
as required. Together with \eqref{lem:AW_v(x/epsilon,epsilon)_V} 
this turns the above Taylor expansion into
\[
\left(\begin{matrix} r \\
m \end{matrix}\right)(y_0,\gamma) 
=
\left(\begin{matrix} \bar{r} \\ 
\bar{m} \end{matrix}\right) 
+ J \epsilon_{\gamma}^{1/4} B \Psi\left(\frac{\sqrt{7}}{4}
\ln\left(\epsilon_{\gamma}\right)\right) \cdot 
\left(b_2+b_{\epsilon_{\gamma}}\right) + o(\epsilon_{\gamma}^{1/4})
\]
for $\gamma \to \infty$,
which up to notation is the desired result, and the proof is complete.
\prfe

\appendix

\setcounter{section}{0}
   \setcounter{subsection}{0}
   \setcounter{subsubsection}{0}
   \setcounter{paragraph}{0}
   \setcounter{subparagraph}{0}
   \setcounter{table}{0}
   \setcounter{equation}{0}
   \setcounter{theorem}{0}
   \renewcommand{\thesection}{\Alph{section}}
   \renewcommand{\theequation}{\Alph{section}.\arabic{equation}}
   \renewcommand{\theproposition}{\Alph{section}.\arabic{proposition}}
   \renewcommand{\thedefinition}{\Alph{section}.\arabic{definition}}
   \renewcommand{\thetheorem}{\Alph{section}.\arabic{theorem}}
   \renewcommand{\thelemma}{\Alph{section}.\arabic{lemma}}

\section{Proof of the linearization result}\label{prooftrafo}
In order to prove Theorem~\ref{RedLem} we first establish 
three auxiliary results.
\begin{lemma} \label{RedLem_Lemma1}
Assume that $a_0$, $A$, and $a_2$ are as in Theorem~\ref{RedLem},
satisfying the conditions
\eqref{RedLem_C0}, \eqref{RedLem_C1}, and \eqref{RedLem_C2}.
Then there exist $\bar{t}\in \,]0,T]$ and a function
$\gamma_0 \in C([0,\bar{t}\,];\R^2)\cap C^1(]0,\bar{t}\,];\R^2)$ 
which satisfies the estimate
\be \label{RedLem_Lemma1_LsgSchranke}
|\gamma_0(t)| 
\leq C t^{-\mu}\int_0^t \sigma^{\mu}\omega(\sigma)\frac{d\sigma}{\sigma},
\ 0\leq t \leq \bar{t},
\ee
and solves the differential equation \eqref{RedLem_u(t)} on $]0,\bar{t}\,]$.
\end{lemma}
{\bf Proof.}
First we note that the right hand side in \eqref{RedLem_Lemma1_LsgSchranke}
converges to $0$ as $t\to 0$.
We define $\tilde{A}(t) := A(t) - A(0)$ and 
$\tilde{a}(t, u) := a_0(t) - \tilde{A}(t) \, u + a_2(t,u)$. 
Then the differential equation~\eqref{RedLem_u(t)} takes the form
\be \label{RedLem_Lemma1_u(t)}
t\frac{du}{dt} = - A(0)\, u + \tilde{a}(t,u).
\ee
Motivated by variation of constants we establish
a solution $\varphi$ of \eqref{RedLem_Lemma1_u(t)} on 
$\left]0, \bar{t} \,\right]$ with $\varphi(t)\to 0$ as 
$t\to 0$ by establishing a solution to
the integral equation 
\be \label{RedLem_Lemma1_Integralgleichung}
\varphi(t) = T\varphi(t) 
:= \int_0^t \left(\frac{t}{\sigma}\right)^{-A(0)}\, 
\tilde{a}(\sigma, \varphi(\sigma))\frac{d\sigma}{\sigma},
\ 0\leq t \leq \bar{t}.
\ee
To this end, we consider the operator
$T$ as defined on the set
\[
\mathcal F_{\Theta, \bar{t}} := 
\left\{\varphi\in C([0,\bar{t}\,]) \mid
|\varphi(t)|\leq \Theta t^{-\mu} 
\int_0^t \sigma^\mu \omega(\sigma)\frac{d\sigma}{\sigma},\ 
0\leq t\leq \bar{t} \right\},
\]
$\Theta>0$ for the moment being arbitrary, and 
$\bar{t}>0$ sufficiently small so that 
$\Theta \omega(\bar{t})/\mu < \delta$.
Now let $\varphi\in \mathcal F_{\Theta, \bar{t}}$ so that in particular, 
$|\varphi|<\frac{\Theta}{\mu}\omega$. By assumption,
\[
|a_0(t)| \leq \omega(t),\ |\tilde{A}(t)| \leq \omega(t),\ 
|a_2(t,u)| \leq C|u|^2
\]
for all $0\leq t \leq \bar{t}$ and $|u| < \delta$, 
where $C>0$ is independent of $\bar{t}$ and $\Theta$. Then
\begin{eqnarray*}
|T\varphi(t)| 
&\leq& 
C \int_0^t \left(\frac{\sigma}{t}\right)^\mu 
\left[\omega(\sigma)\left(1+|\varphi(\sigma)|\right) 
+ |\varphi(\sigma)|^2\,\right] \frac{d\sigma}{\sigma} \nonumber \\
&\leq& 
C t^{-\mu}\int_0^t \sigma^\mu 
\left[\omega(\sigma)\left(1+\left(1+\frac{\Theta}{\mu}\right)
|\varphi(\sigma)|\right)\right] \frac{d\sigma}{\sigma} \nonumber \\
&\leq& 
C \left[1+\left(1+\frac{\Theta}{\mu}\right)\frac{\Theta}{\mu}
\omega\left(\bar{t}\,\right)\right] t^{-\mu}
\int_0^t \sigma^\mu \omega(\sigma)\frac{d\sigma}{\sigma} \nonumber \\
&\leq& 
\Theta t^{-\mu}\int_0^t \sigma^\mu \omega(\sigma)\frac{d\sigma}{\sigma},
\ 0\leq t\leq \bar{t}\,,
\end{eqnarray*}
i.e., $T\phi \in \mathcal F_{\Theta, \bar{t}}$,
where we can choose $\Theta=2 C$ and decrease $\bar{t}>0$ further
to make the $\Theta$-dependent term in brackets less than $2$; 
notice that 
$\omega$ is continuous and increasing with $\omega(0)=0$. 
Clearly, $\mathcal F_{\Theta, \bar{t}} \subset C([0,\bar{t}\,])$ 
is non-empty, bounded, closed, and convex.
Similarly to the analysis of the operator
$T_{\epsilon}$ in the proof of Lemma~\ref{LemAZ1} one can show that
$T\colon \mathcal F_{\Theta, \bar{t}} \to \mathcal F_{\Theta, \bar{t}}$ 
is compact and by Schauder's theorem has a fixpoint 
$\gamma_0\in\mathcal F_{\Theta, \bar{t}}$. As a solution 
to~\eqref{RedLem_Lemma1_Integralgleichung} the function $\gamma_0$ 
is continuously differentiable on $]0,\bar{t}\,]$ and 
satisfies \eqref{RedLem_u(t)} there. 
The estimate \eqref{RedLem_Lemma1_LsgSchranke} follows from the
definition of the set $\mathcal F_{\Theta, \bar{t}}$, and
the proof is complete.
\prfe

\begin{lemma} \label{RedLem_Lemma2}
Consider a matrix-valued function $A$ with the properties
specified in Theorem~\ref{RedLem} so that in particular
the condition~\eqref{RedLem_C1}
holds. Then there exist $0<\bar{t}\leq T$,
a function 
$\Gamma \in C([0,\bar{t}\,];\R^{2\times 2})
\cap C^1(]0,\bar{t}\,];\R^{2\times 2})$, and a constant $C>0$ such that
\[
|\Gamma(t)-E|  \leq  
C\int_0^t \omega(\sigma)\frac{d\sigma}{\sigma},\ 0\leq t\leq \bar{t},
\]
and
\be \label{RedLem_Lemma2_Gamma(t)}
t\frac{d\Gamma}{dt}  =  -A(t)\, \Gamma +\Gamma\, A(0)
\ \mbox{on}\ ]0,\bar{t}\,].
\ee
\end{lemma}
{\bf Proof.}
If $\Gamma$ is a complex-valued solution to \eqref{RedLem_Lemma2_Gamma(t)}
with $\Gamma(0)=\mathrm{id}$, then its real part has the same properties.
Hence it is sufficient to determine such a complex
valued solution under the assumption that
$A(0)$ is diagonal with eigenvalues 
$\lambda_{1/2} = \mu \pm \nu i$, $\mu>0$.
Let
\be \label{RedLem_Lemma2_AGamma}
A(t)  =  \left(\begin{matrix} \lambda_1 + a_{11}(t) & a_{12}(t) \\ 
a_{21}(t) & \lambda_2 + a_{22}(t) \end{matrix}\right),\ 
\mbox{and}\
\Gamma  =  \left(\begin{matrix} \gamma_{11} & \gamma_{12} \\ 
\gamma_{21} & \gamma_{22} \end{matrix}\right)\,,
\ee
in particular by \eqref{RedLem_C1}, 
$|a_{ij}(t)| < \omega(t)$ for $i,j=1,2$ and 
$0\leq t \leq T$.
Then  \eqref{RedLem_Lemma2_Gamma(t)} takes the form
\begin{eqnarray}
t\frac{d\gamma_{11}}{dt} 
&=& 
- a_{11}(t)\gamma_{11} - a_{12}(t) \gamma_{21},
\label{RedLem_Lemma2_Gamma(t)_11}\\
t\frac{d\gamma_{12}}{dt} 
&=& 
-(2\nu i + a_{11}(t))\gamma_{12} - a_{12}(t) \gamma_{22},  
\label{RedLem_Lemma2_Gamma(t)_12}\\
t\frac{d\gamma_{21}}{dt} 
&=& 
- a_{21}(t) \gamma_{11} + (2\nu i - a_{22}(t))\gamma_{21},
\label{RedLem_Lemma2_Gamma(t)_21}\\
t\frac{d\gamma_{22}}{dt} 
&=& 
- a_{21}(t)\gamma_{12} - a_{22}(t) \gamma_{22}.           
\label{RedLem_Lemma2_Gamma(t)_22}
\end{eqnarray}
Since we require that $\gamma_{11}(0)=1$ 
we assume for the moment that $\gamma_{11}\neq 0$.
By \eqref{RedLem_Lemma2_Gamma(t)_11} and \eqref{RedLem_Lemma2_Gamma(t)_21} 
the function $\varphi := \gamma_{21}/\gamma_{11}$ must satisfy
the equation
\[
t\frac{d\varphi}{dt} =
(2\nu i + a_{11}(t) - a_{22}(t) )\varphi  
-a_{21}(t) + a_{12}(t)\varphi^2
=: b_0(t)\varphi + b_1(t,\varphi)
\]
for which we need a solution with  $\varphi(0)=0$. 
By variation of constants this is turned into
the integral equation
\be \label{RedLem_Lemma2_Tvarphi}
\varphi(t)  =  T\varphi(t) 
:= \int_0^t \exp\left(\int_\sigma^t b_0(\tau)\frac{d\tau}{\tau}\right) 
b_1(\sigma, \varphi(\sigma)) \frac{d\sigma}{\sigma}.
\ee
Consider the set  $\mathcal F_{\Theta, \bar{t}}$ 
of continuous functions $\varphi$ on $[0,\bar{t}\,]$ 
which satisfy the estimate
\[
|\varphi(t)| \leq 
\Theta \int_0^t \omega(\sigma)\frac{d\sigma}{\sigma},\
0\leq t \leq \bar{t}.
\] 
For such a function,
\begin{eqnarray} \label{RedLem_Lemma2_absTvarphi}
|T\varphi(t)| 
&\leq& 
\int_0^t \exp\left(2 \int_0^t \omega(\tau)\frac{d\tau}{\tau}\right) 
\omega(\sigma)\left(1+\left(\Theta 
\int_0^t \omega(\tau)\frac{d\tau}{\tau}\right)^2\right) 
\frac{d\sigma}{\sigma} \nonumber\\
&<& 
\Theta \int_0^t \omega(\sigma)\frac{d\sigma}{\sigma},\ 0\leq t \leq \bar{t},
\end{eqnarray}
provided $\Theta>0$ is sufficiently large and $\bar{t}>0$ 
is sufficiently small.
As before, Schauder's theorem applies to the map 
$T: \mathcal F_{\Theta, \bar{t}}\to \mathcal F_{\Theta, \bar{t}}$ 
so that a solution $\varphi\in\mathcal F_{\Theta, \bar{t}}$ 
to \eqref{RedLem_Lemma2_Tvarphi} exists. Using
this solution,
\eqref{RedLem_Lemma2_Gamma(t)_11} turns into the linear equation
\[
t\frac{d\gamma_{11}}{dt}  =  -(a_{11}(t) + a_{12}(t)\varphi(t))\gamma_{11}
\]
for $\gamma_{11}$, which together with $\gamma_{11}(0)=1$ is solved by
\[ 
\gamma_{11}(t)  =  
\exp\left(-\int_0^t(a_{11}(\sigma) + a_{12}(\sigma)\varphi(\sigma))
\frac{d\sigma}{\sigma}\right),\ 0\leq t \leq\bar{t};
\]
notice that this function is positive everywhere.
From \eqref{RedLem_Lemma2_absTvarphi} it follows that 
\[
|\gamma_{11}(t)-1|, |\gamma_{21}(t)| 
< C \int_0^t \omega(\sigma)\frac{d\sigma}{\sigma}.
\]
Analogous arguments applied to \eqref{RedLem_Lemma2_Gamma(t)_12} and 
\eqref{RedLem_Lemma2_Gamma(t)_22}  yield $\gamma_{12}$ and $\gamma_{22}$.
\prfe
\begin{lemma} \label{RedLem_Lemma3}
Consider a function $a_2$ with the properties
specified in Theorem~\ref{RedLem} so that in particular
the condition~\eqref{RedLem_C2} holds.
Then there exist $0<\tilde{\delta}<\delta$ and a function 
$\gamma_2=\gamma_2(t,t_0,z_0)$ defined on the set 
$\{(t,t_0,z_0)\in \,]0,T]^2\times B_{\tilde{\delta}}(0) \mid t\geq t_0\}$ 
such that the following holds:
For all data $t_0\in \,]0,T[$ and $z_0 \in B_{\tilde{\delta}}(0)$
and with $z(t)= (t/t_0)^{- A(0)}z_0$, 
the definition $u := z + \gamma_2(\cdot,t_0,z_0)$ yields a solution
of the initial value problem
\be \label{RedLem_Lemma3_u(t)}
t\frac{du}{dt}  =  -A(0)\, u + a_2(t,u),\ u(t_0) = z_0,
\ee
which exists on the interval $[t_0,T]$. 
Moreover, there exists a constant $C>0$ which is independent
of $t_0$ and $z_0$ such that
\[ 
|\gamma_2(t,t_0,z_0)|  \leq  C (t/t_0)^{-\mu}|z_0|^2 ,\ t\in[t_0,T] .
\]
For $t_0\in \,]0,T]$ and $z_0\in B_{\tilde{\delta}}(0)$,
$\gamma_2(t_0,t_0,z_0)  =  0$.
\end{lemma}
{\bf Proof.}
Let $0<\tilde \delta < \delta$ for the moment be arbitrary.
The initial value problem \eqref{RedLem_Lemma3_u(t)} has 
a unique solution $u$ which exists
on some interval $I_0\subset [t_0,T]$ 
with $t_0\in I_0$ and satisfies the estimate
$|u(t)| < \delta$ there. We choose this interval
maximal and want to show that for $\tilde\delta$ sufficiently small
it equals $[t_0,T]$.
By variation of constants,
\be \label{RedLem_Lemma3_VdK_U}
u(t)  =  z(t)
+ \int_{t_0}^t \left(\frac{t}{\sigma}\right)^{-A(0)}\!\!\, 
a_2(\sigma, u(\sigma))\frac{d\sigma}{\sigma},
\ee
for $t\in I_0$. With $C>0$ denoting the larger of the
two constants from \eqref{tBest} and \eqref{RedLem_C2}, this yields the estimate
\begin{eqnarray*} 
|u(t)| 
&\leq& 
C\left(\frac{t_0}{t}\right)^\mu|z_0| 
+ C\int_{t_0}^t \left(\frac{\sigma}{t}\right)^\mu 
|u(\sigma)|^2\frac{d\sigma}{\sigma}\\
&\leq&
C\left(\frac{t_0}{t}\right)^\mu|z_0| 
+ \mu \int_{t_0}^t \left(\frac{\sigma}{t}\right)^\mu 
|u(\sigma)| \frac{d\sigma}{\sigma},
\end{eqnarray*}
where we require in addition that $\tilde \delta < \mu/C$,
and the last estimate holds as long as $|u(\sigma)|< \mu/C$
with the definition of $I_0$ adjusted accordingly.
We apply Gronwall's lemma to the function
$t^\mu |u(t)|$ to find that
\be \label{RedLem_Lemma3_Gronwall_U}
|u(t)| \leq t^{-\mu} C t_0^\mu |z_0| 
\left(t /t_0\right)^\mu = C |z_0| < C \tilde \delta 
\ee
for $t\in I_0$. So if we choose
$0<\tilde{\delta} < \min\{\delta, \delta/C,\mu/C^2\}$,
then $z_0\in B_{\tilde{\delta}}(0)$ implies that $I_0=[t_0,T]$.

The first term in \eqref{RedLem_Lemma3_VdK_U} has the desired form. 
We define
\[
\gamma_2(t,t_0,z_0) := u(t) - z(t),\ 
\xi := \gamma_2/|z| = (u-z)/|z|,
\]	
provided $z_0\neq 0$ which implies that $z\neq 0$ on $]0,\infty[$.
Since $\xi(0)=0$ it follows that $|\xi|< 1$ on some
maximally chosen interval $I\subset [t_0,T]$ with $t_0\in I$. 
We need to estimate $\xi$ against $|z_0|$, and in doing so we assume without
loss of generality that $A(0)$ is diagonal with its
eigenvalues $\mu\pm i\nu$ on the diagonal. The function
$\xi$ satisfies the differential equation
\begin{eqnarray} 
t\frac{d\xi}{dt} 
&=&
\frac{1}{|z|}\left(t\frac{du}{dt}-t\frac{dz}{dt}\right) 
- \frac{1}{2|z|^3}
\left(t\frac{dz}{dt} \cdot \bar{z} 
+ z\cdot t\frac{d\bar{z}}{dt}\right) \gamma_2 
\nonumber\\
&=& 
\frac{1}{|z|}\left( -A(0) u + a_2(t,u) + A(0) z\right)
- \frac{1}{|z|^2}
\mathrm{Re}\left( (-A(0) z)\cdot \bar{z} \right) \xi
\nonumber\\
&=&  
\left(-A(0) + \mu E\right) \xi + \frac{1}{|z|} a_2(t,u)\nonumber\\
&=&  
\left(\begin{matrix} -i\nu & 0\\ 0 & i\nu\end{matrix}\right) \xi 
+ \frac{1}{|z|}a_2(t,u).
\end{eqnarray}
By variation of constants the function $\xi$ satisfies the relation
\[
\xi(t)  =  \int_{t_0}^t 
\left(\begin{matrix} \left(t/\sigma\right)^{-i\nu} & 0 \\ 
0 & \left(t/\sigma\right)^{i\nu}\end{matrix}\right)
\,\frac{1}{|z(\sigma)|}a(\sigma, (z+|z|\xi)(\sigma))\frac{d\sigma}{\sigma}.
\]
Since we already know that $|u| < \delta$ and also
$|u| < 2 |z|$ as long as $|\xi| < 1$, it follows that
\begin{eqnarray} \label{RedLem_Lemma3_|xi|}
|\xi(t)| 
&\leq& 
\int_{t_0}^t \frac{C}{|z(\sigma)|}
\left|(z+|z|\xi)(\sigma)\right|^2\frac{d\sigma}{\sigma} 
\leq 
C \int_{t_0}^t |z(\sigma)|\,\frac{d\sigma}{\sigma} \nonumber\\
&\leq& 
C |z_0| \int_{t_0}^t \left(\frac{\sigma}{t_0}\right)^{-\mu}
\frac{d\sigma}{\sigma} 
= \frac{C}{\mu} |z_0|\, t_0^{\mu} \left(t_0^{-\mu} -t^{-\mu}\right)
\leq
C^\ast |z_0|\quad
\end{eqnarray}
on $I$, where $C^\ast >0$ is independent of $t_0$ and $z_0$. 
We decrease $\tilde \delta$
one final time so that $\tilde{\delta} < 1/C^\ast$. Then
the last estimate implies that for $z_0\in B_{\tilde{\delta}}(0)$ 
the estimate $|\xi(t)| < C^\ast \tilde\delta < 1$ holds on $I$,
which by definition of that
interval implies that $I=[t_0,T]$. 
In view of the definition of $\xi$ and the estimates
\eqref{RedLem_Lemma3_|xi|} and $|z(t)|\leq C (t_0/t)^{\mu} |z_0|$ 
the proof is complete.
\prfe

\noindent
{\bf Proof of Theorem~\ref{RedLem}.}
By Lemma~\ref{RedLem_Lemma1} there exists for $0<\bar{t}<T$
sufficiently small a solution $\gamma_0$ to \eqref{RedLem_u(t)}
on the interval $]0,\bar{t}\,]$. 
Now assume that $\Delta$ solves the equation 
\be \label{RedLem_Delta(t)}
t\frac{d\Delta}{dt} = -B(t)\, \Delta + b_2(t,\Delta)
\ee
on some interval $I\subset~]0,\bar{t}\,]$, where
for $0\leq t\leq \bar{t}$,
\begin{eqnarray*}
B(t) 
&:=& 
A(t) - \partial_u a_2(t,\gamma_0(t)), \\
b_2(t,\Delta) 
&:=& 
a_2(t,\gamma_0(t)+\Delta) - a_2(t,\gamma_0(t)) 
- \partial_u a_2(t,\gamma_0(t))\, \Delta.
\end{eqnarray*}
Then $u= \gamma_0 + \Delta$ is a solution to \eqref{RedLem_u(t)} 
on the interval $I$:
\begin{eqnarray*}
t\frac{du}{dt} 
&=& 
a_0(t) - A(t)\, \gamma_0(t) + a_2(t, \gamma_0(t)) \\
&&
{}-\left(A(t)-\partial_u a_2(t,\gamma_0(t))\right) 
\left(u-\gamma_0(t)\right) \\
&&
{} + a_2(t,u) - a_2(t,\gamma_0(t)) - 
\partial_u a_2\left(t,\gamma_0(t)\right)\, (u-\gamma_0(t))\\
&=& 
a_0(t) - A(t)\, u + a_2(t,u).
\end{eqnarray*}
The assumptions on $a_2$ in Theorem~\ref{RedLem} imply that
$|\partial_u a_2(t,\gamma_0(t))|\leq C|\gamma_0(t)|$ so that
$B$ satisfies the condition~\eqref{RedLem_C1}; notice that $B(0)=A(0)$. 
Let $\Gamma$ correspond to $B$ as provided by Lemma~\ref{RedLem_Lemma2}.
If $\bar{t}>0$ is chosen sufficiently small, then $\Gamma(t)$ is
invertible for $t\in [0,\bar{t}\,]$. 

Assume now that $\tilde{\Delta}$ solves the equation 
\be \label{RedLem_TildeDelta(t)}
t\frac{d\tilde{\Delta}}{dt} = -A(0)\, \tilde{\Delta} 
+ \Gamma^{-1}(t)\, b_2(t,\Gamma(t)\tilde{\Delta})
\ee
on some	interval $I\subset ]0,\bar{t}\,]$. Then
$\Delta := \Gamma\,\tilde{\Delta}$ solves  
\eqref{RedLem_Delta(t)} on $I$:
\begin{eqnarray*}
t\frac{d\Delta}{dt} 
&=& 
t\frac{d\Gamma}{dt}\,\tilde{\Delta} 
+ \Gamma\, t\frac{d\tilde{\Delta}}{dt} \\
&=& 
\left(-B(t)\,\Gamma + \Gamma\, B(0)\right) \tilde{\Delta}
+ \Gamma\, \left(-A(0)\, \tilde{\Delta} 
+ \Gamma^{-1}(t)\, b_2(t,\Gamma(t)\tilde{\Delta})\right).
\end{eqnarray*} 
By Taylor expansion we see that the function
$c_2(t,\tilde\Delta):=\Gamma^{-1}(t)\, b_2(t,\Gamma(t)\tilde{\Delta})$ 
in \eqref{RedLem_TildeDelta(t)} satisfies the condition~\eqref{RedLem_C2}
for $t\in [0,\bar{t}\,]$ and $\tilde\Delta$ in a suitable ball about the origin.
Hence Lemma~\ref{RedLem_Lemma3} provides a constant
$\tilde\delta>0$ and a function $\gamma_2$ such that for
all data $t_0\in \,]0,\bar{t}[$ and $z_0\in B_{\tilde{\delta}}(0)$ and with
$z(t)=(t/t_0)^{-A(0)}z_0$ the function 
$\tilde{\Delta}(t) = z(t) + \gamma_2(t,t_0, z_0)$ 
solves \eqref{RedLem_TildeDelta(t)} on $[t_0,\bar{t}\,]$. 
This in turn implies that 
$u=\gamma_0 + \Delta = \gamma_0 + \Gamma (z + \gamma_2)$
solves \eqref{RedLem_u(t)},
and the proof is complete.
\prfe

\section{The polytropic case}\label{polyrm}
\setcounter{equation}{0}
To conclude this paper we quickly give the
argument for proving (\ref{poly_RM}), which relies
on the scaling property of polytropic steady states.
For this argument we consider the more general ansatz
\be \label{polytrkl}
f(x,v) = (E_0 -E)_+^k L^l.
\ee
We fix $k$ and $l$ which lead to steady states with finite radius 
and finite mass, and we fix one such state
$(\tilde{f},\tilde{U})$ of finite, positive mass and
denote its cut-off energy by $\tilde{E}_0$. Then the scaling
transformation
\be \label{Polytrope_Trafo_f}
f(x,v) = \beta^{-\frac{4k+2l}{2l+2}}
\tilde{f}\left(\beta^{-\frac{2(k+l)+1}{2+2l}} x, \beta v\right),
\ee
\be \label{Polytrope_Trafo_U}
E_0 - U(x) = \beta^{-2}
\left(\tilde{E}_0 - \tilde{U}\left(\beta^{-\frac{2(k+l)+1}{2+2l}} x\right)\right)
\ee
with $\beta >0$ yields all other polytropic steady states with the
same $k$ and $l$ and finite positive mass and radius. To check this is a 
straight forward computation which crucially relies on the fact
that for such polytropes,
\[
\rho(x) = c \left(E_0-U(x)\right)_+^{k+l+3/2},
\]
which implies that $\rho$ inherits a corresponding scaling law.
In other words, all the members of the one-parameter family of
steady states corresponding to the fixed choice of $k$ and $l$
are obtained from a fixed one by such a scaling transformation.
If we compute how $M$ and $R$ behave under this scaling and
eliminate the scaling parameter $\beta$, the relation
\[
M(R) = C \,R^{\frac{2(k-l)-3}{2(k+l)+1}}
\]
pops out, which for $l=0$ reduces to (\ref{poly_RM}).

\begin{figure} 
\centering
\resizebox*{!}{0.44\textwidth}{\input{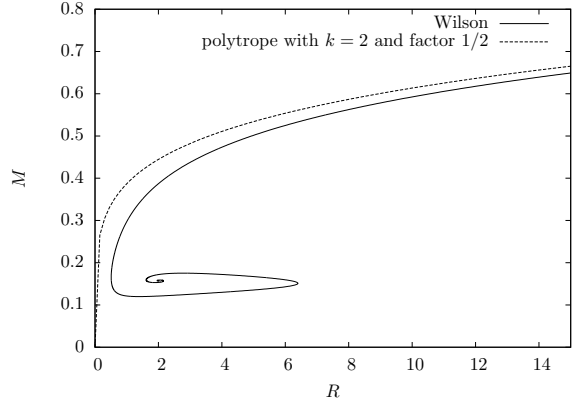}}
\caption{Mass-radius diagram for the Wilson model}
\label{wilson_fig}
\end{figure}

Numerically, we have observed a relation between the Woolley-Dickens,
King, or Wilson models on the one hand and certain polytropes
on the other hand, which we want to illustrate for the 
Wilson case.
For $\eta = E_0-E >0$ small, i.e., close to the cut-off energy,
the Wilson ansatz satisfies the relation
\[
\phi(\eta) = e^\eta -1-\eta = \frac{1}{2} \eta^2 + O(\eta^3),\ \eta \to 0.
\]
In Figure~\ref{wilson_fig} we plot
the numerically computed mass-radius spiral
for the Wilson model and the exact curve for the polytrope with $k=2$
and a factor $1/2$.
We see that the mass-radius spiral for the Wilson model lies completely below
the polytropic curve. In particular, this indicates that for the Wilson model
mass and radius are finite for all $\gamma>0$.
The same relation is numerically observed between the Woolley-Dickens
or King model and the corresponding polytropes with $k=0$ or $k=1$, respectively,
but a rigorous investigation of this relation must be postponed to 
future research.


\begin{thebibliography}{10}

\bibitem{AR1}
{\sc Andr\'{e}asson, H., Rein, G.},
A numerical investigation of the stability of steady states 
and critical phenomena for the spherically symmetric 
Einstein-Vlasov system.
{\em Class.\ Quantum Grav.} {\bf 23}, 3659--3677 (2006).

\bibitem{AR2}
{\sc Andr\'{e}asson, H., Rein, G.},
On the steady states of the spherically symmetric
Einstein-Vlasov system.
{\em Class.\ Quantum Grav.}\ {\bf 24}, 1809--1832 (2007).

\bibitem{BFH}
{\sc Batt, J., Faltenbacher, W., Horst, E.},
Stationary spherically symmetric models in stellar dynamics.
{\em Arch.\ Rational Mech.\ Anal.}\ {\bf 93}, 159--183  (1986).

\bibitem{BT}
{\sc Binney, J., Tremaine, S.},
{\em Galactic Dynamics}, Princeton University Press, Princeton 1987.

\bibitem{GNN}
{\sc Gidas, B., Ni, W.-M., Nirenberg, L.},
Symmetry and related properties via the maximum principle.
{\em Commun.\ Math.\ Phys.}\ {\bf 68}, 209--243 (1979).

\bibitem{GR}
{\sc Guo, Y., Rein, G.},
A non-variational approach to nonlinear stability in stellar
dynamics applied to the King model.
{\em Commun.\ Math.\ Phys.}
{\bf 271}, 489--509 (2007).

\bibitem{HR1}
{\sc Had\v{z}i\'{c}, M., Rein, G.},
Stability for the spherically symmetric Einstein-Vlasov 
system---a coercivity estimate.
{\em Math.\ Proc.\ Camb.\ Phil.\ Soc.}\
{\bf 155}, 529--556 (2013).

\bibitem{HR2}
{\sc Had\v{z}i\'{c}, M., Rein, G.},
On the small redshift limit of steady states of the spherically 
symmetric Einstein-Vlasov system
and their stability.
{\em Math.\ Proc.\ Camb.\ Phil.\ Soc.}\
{\bf 159}, 529--546 (2015).

\bibitem{HRU}
{\sc Heinzle, J., Rendall, A., Uggla, C.},
Theory of Newtonian self-gravitating stationary
spherically symmetric systems.
{\em Math.\ Proc.\ Camb.\ Phil.\ Soc.}\ {\bf 140}, 177--192 (2006).

\bibitem{LMR}
{\sc Lemou, M., M\'{e}hats, F., Rapha\"el, P.},
Orbital stability of spherical galactic models.
{\em Invent.\ math.}\ {\bf 187}, 145--194 (2012).

\bibitem{Mak}
{\sc Makino, T.},
On the spiral structure of the $(R,M)$-diagram for a
stellar model of the Tolman-Oppenheimer-Volkoff equation.
{\em Funkcialaj Ekvacioj} {\bf 43}, 471--489  (2000).

\bibitem{Ramming}
{\sc Ramming, T.},
{\em \"Uber Familien sph\"arisch symmetrischer station\"arer
L\"o\-sun\-gen des Vlasov-Poisson-Systems},
PhD thesis, Bayreuth 2012.

\bibitem{RammRein}
{\sc Ramming, T., Rein, G.},
Spherically symmetric equilibria for
self-gravitating kinetic or fluid models in the non-relativistic
and relativistic case---A simple proof for finite extension.
{\em SIAM J.\ on Mathematical Analysis}
{\bf 45}, 900--914 (2013).

\bibitem{Rein02}
{\sc Rein, G.},
Reduction and a concentration-compactness principle
for energy-Casimir functionals.
{\em SIAM J.\ Math.\ Anal.}\
{\bf 33}, 896--912 (2002).

\bibitem{Rein03}
{\sc Rein, G.},
Non-linear stability of gaseous stars.
{\em Arch. Rational Mech. Anal.}\
{\bf 168}, 115--130 (2003).

\bibitem{Rein07}
{\sc Rein, G.},
Collisionless kinetic equations from astrophysics---The Vla\-sov-Poisson
system.
In {\em Handbook of Differential Equations, Evolutionary Equations,
vol.~3}, ed.\ by C.~M.~Dafermos and E.~Feireisl,
Elsevier (2007).

\bibitem{Rein14}
{\sc Rein, G.},
Galactic dynamics in MOND---Existence of equilibria
with finite mass and compact support.
{\em Kinetic and Related Models}
{\bf 8}, 381--394 (2015)


\bibitem{RR00}
{\sc Rein, G., Rendall, A.},
Compact support of spherically symmetric equilibria in
non-relativistic and relativistic galactic dynamics.
{\em Math.\ Proc.\ Camb.\ Phil.\ Soc.}\
{\bf 128}, 363--380 (2000).

\bibitem{Sans}
{\sc Sansone, G.},
Sulle soluzione die Emden dell'equazione di Fowler.
{\em Rend.\ Mat.\ Roma} {\bf 1}, 163--176 (1940).

\end{thebibliography}
\end{document}